\documentclass{emulateapj}
\usepackage{natbib}
\usepackage{multirow}
\shorttitle{Kinematic Modeling of NGC 3044 and NGC 4302}
\shortauthors{Zschaechner, Rand, \& Walterbos}

\begin{document}

\title{Investigating Disk-Halo Flows and Accretion:  A Kinematic and Morphological Analysis of Extraplanar \sc{HI} in NGC 3044 and NGC 4302} 
\author{\sc Laura K. Zschaechner\altaffilmark{1,2}, Richard J. Rand\altaffilmark{2}, {\sc and} Rene Walterbos\altaffilmark{3}}
\altaffiltext{1}{Max Planck Institute for Astronomy - K\"{o}nigstuhl 17, 69117 Heidelberg - Germany; zschaechner@mpia.de}
\altaffiltext{2}{Department of Physics and Astronomy, University of New Mexico, 1919 Lomas Blvd NE, Albuquerque, New Mexico 87131 - USA; rjr@phys.unm.edu}
\altaffiltext{3}{Department of Astronomy, New Mexico State University, PO Box 30001, MSC 4500, Las Cruces, New Mexico 88003 - USA; rwalterb@nmsu.edu}
\slugcomment{Accepted to Astrophysical Journal on 10 November 2014}
\begin{abstract}

\par
To further understand the origins of and physical processes operating in extra-planar gas, we present observations and kinematic models of {\sc H\,i} in the two nearby, edge-on spiral galaxies NGC 3044 and NGC 4302.  We model NGC 3044 as a single, thick disk.  Substantial amounts of extra-planar {\sc H\,i} are also detected.  We detect a decrease in rotation speed with height (a lag) that shallows radially, reaching zero at approximately R$_{25}$.  The large-scale kinematic asymmetry of the approaching and receding halves suggests a recent disturbance.  The kinematics and morphology of NGC 4302, a Virgo Cluster member, are greatly disturbed.  We model NGC 4302 as a combination of a thin disk and a second, thicker disk, the latter having a hole near the center.  We detect lagging extra-planar gas, with indications of shallowing in the receding half, although its characteristics are difficult to constrain.  A bridge is detected between NGC 4302 and its companion, NGC 4298.  We explore trends involving the extra-planar {\sc H\,i} kinematics of these galaxies, as well as galaxies throughout the literature, as well as possible connections between lag properties with star formation and environment.  Measured lags are found to be significantly steeper than those modeled by purely ballistic effects, indicating additional factors.  Radial shallowing of extra-planar lags is typical and occurs between 0.5$R_{25}$ and $R_{25}$, suggesting internal processes are important in dictating extra-planar kinematics.

\end{abstract}


\section{Introduction}\label{introduction}

\par
    Provided galactic star formation rates (SFRs) are relatively constant, if there is no source of new material to fuel star formation (SF) in galaxies as they evolve, their gas reservoirs are depleted (Putman et al. 2012 and references therein).  Such a scenario would result in little ongoing SF in most galaxies, which is observationally not the case.  Furthermore, without a source of fresh material, newly-formed stars and the interstellar medium (ISM) would have higher metallicities than those observed (e.g.\ \citealt{2008NewA...13..314C}, \citealt{2003ApJ...596...47S}). Accretion of primordial gas from external sources could solve this dilemma.  Thus, it is desirable to gauge the ongoing accretion rate for a large sample of galaxies (see $\S$~\ref{theory}).

\par
   Extra-planar layers act as the interface between the intergalactic medium (IGM) and the main disk of the galaxy.  Therefore, any accreted material must pass through this zone before it may be used to fuel SF within the disk.   Only recently have extra-planar layers been studied in detail, with an even more recent emphasis on their kinematics (e.g.\ Heald et al. 2007, Oosterloo et al. 2007), in order to constrain whether their origins are primarily internal, external, or both.

\par
   Extra-planar material is observed ubiquitously in spiral galaxies (e.g.\ \citealt{2012ARA&A..50..491P} and references therein).  Notable extra-planar components include hot gas (e.g.\ \citealt{2006A&A...448...43T}, \citealt{2013ApJ...762...12H}), relativistic particles (e.g.\ \citealt{1999AJ....117.2102I}), dust (e.g.\ \citealt{1999AJ....117.2077H}), diffuse ionized gas (DIG) (e.g.\ \citealt{1996ApJ...462..712R}; \citealt{2003A&A...406..493R}), and {\sc H\,i} (e.g.\ \citealt{1997ApJ...491..140S}, \citealt{2007AJ....134.1019O}).  With the exception of {\sc H\,i}, a connection has been made between each extra-planar component and SF within the disk on both global and localized scales, indicating that they are largely due to disk-halo flows.  One may assume that extra-planar {\sc H\,i} may also be attributed to disk-halo flows as demonstrated in multiple studies including \citet{1994ApJ...429..618I} and \citet{2005A&A...431...65B}, but there are indications of {\sc H\,i} accretion [e.g.\ HVCs in the Milky Way \citep{1997ARA&A..35..217W},  counterrotating clouds in NGC 891 seen by \citealt{2007AJ....134.1019O}], rendering the contributions of each source unknown.

\par
  The kinematics of extra-planar material is also of use in determining its origins.  Of recent interest is the discovery of negative gradients in rotation speed with height above the midplane (lags e.g.\ \citealt{2007AJ....134.1019O}, \citealt{2007ApJ...663..933H}).  Some lag is expected due to conservation of angular momentum in a disk-halo flow - gas is ejected from the midplane and thus, moving further away from the center of the gravitational potential, migrates to a larger radius, and then must slow to conserve angular momentum before falling back to the disk.  However, if the observed lag differs from that predicted by such gravitational effects (considered in greater depth in ballistic models of disk-halo flow by \citealt{2002ApJ...578...98C}, \citealt{2006MNRAS.366..449F}), then other factors must also contribute.  Within galaxies, possibilities include extra-planar radial pressure gradients and magnetic tension (Benjamin 2002, 2012), but too little is known of either of these to make meaningful theoretical constraints at this time, thus additional observations and empirical models of a large number of galaxies are necessary.  

\par
   Recent theoretical simulations have explored lag kinematics in terms of disk-halo flows and accretion.  Simulations presented by \citet{2011MNRAS.415.1534M} rely on momentum and heat exchanges occurring in disk-halo flows between cooler clouds launched from the disk and a hot corona.  During these exchanges, gas is stripped from a cloud by the corona.  The stripped gas streams behind the cloud, mixing with the hot coronal gas, resulting in relatively efficient cooling of the coronal gas compared to the otherwise lengthy cooling times for such hot gas as described in \citet{2008MNRAS.386..935F}.  The trailing, mixed gas condenses into smaller {\sc H\,i} clouds, and then falls toward the disk.   \citet{2011MNRAS.415.1534M} find that the momentum loss to the corona would contribute approximately $-$7 km s$^{-1}$ kpc$^{-1}$ to the deceleration of clouds (approximately 50$\%$ of the total $-$15 km s$^{-1}$ kpc$^{-1}$ observed in NGC 891).  When combined with ballistic effects, this model would help to reconcile the discrepancies between aforementioned purely ballistic models and observations.  However, these simulations focus on NGC 891 and the Milky Way, and lags have been shown to differ substantially among galaxies (e.g.\ \citealt{2007ApJ...663..933H}, \citealt{2008A&ARv..15..189S}, \citealt{2012ApJ...760...37Z}, \citealt{2013MNRAS.434.2069K}).

\par
   Within the past few years, in a few cases observations and models have shown that lags shallow with radius (\citealt{2007AJ....134.1019O}, Zschaechner et al. 2011, 2012) although it is not yet clear if this is a general result.   Based on conservation of angular momentum, some shallowing is expected - material at smaller radii experiences larger relative changes in the gravitational potential when ejected from the midplane to a given height (and thus a greater lag) than material at large radii. Given the lack of agreement already found for lag magnitudes, it is likely that additional factors must also affect lag properties with radius.  Thus, the nature of the radial variations can also help to constrain the possibilities discussed above.

\par
   Recent observations of several nearby edge-on spiral galaxies have aided greatly in increasing our understanding of extra-planar {\sc H\,i}, but it is still too early to search for definite trends concerning its kinematic and morphological properties and how they may relate to star formation or accretion.  The HALOGAS survey \citep{2011A&A...526A.118H} has been instrumental in providing a consistently observed and modeled sample.  However, the HALOGAS sample is largely comprised of galaxies having moderate SFRs, omitting vital extrema if the relation between extra-planar {\sc H\,i} and SF is to be investigated.  Thus, we supplement the HALOGAS sample with Karl G. Jansky Very Large Array (VLA) observations of NGC 3044, which has a moderate global SFR, but high $L_{TIR}$/${D_{25}}^2$ (i.e. the total IR luminosity divided by the optical area, which is an indicator of star formation per unit area and is the more relevant quantity for this work).  This quantity is much lower for the edge-on galaxies in the HALOGAS sample.

\par
    We also include VLA observations of NGC 4302 for three main reasons:  1)  NGC 4302 has a measured DIG lag \citep{2007ApJ...663..933H}.  By comparing the {\sc H\,i} and DIG lags, we may consider the likelihood that they have similar origins.  2)  The $L_{TIR}$/${D_{25}}^2$ is also quite high compared to the HALOGAS sample.  3) NGC 4302 is a member of the Virgo Cluster with a nearby companion galaxy, thus adding variety to the types of environments considered.

\subsection{NGC 3044}

\par
\begin{figure}
\begin{center}
\includegraphics[width=80mm]{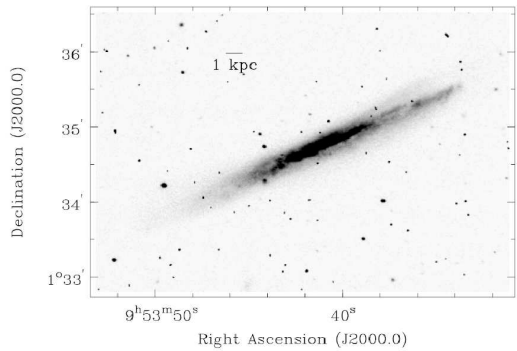}
\caption[NGC 3044 in Red Continuum]{\scriptsize\textit{Figure 2a from \citet{2000ApJ...536..645C} shows NGC 3044 in red continuum.  Note the asymmetries in the disk.  The faint extension on the NW side is particularly suggestive of an interaction.} \label{n3044redcont}}
\end{center}
\end{figure}

      Basic observed parameters of NGC 3044 are given in Table~\ref{n3044tbl-1}.  By examining Figure~\ref{n3044redcont}, note how the disk appears to be asymmetric and disturbed, possibly through a recent interaction or merger.  \citet{1988ApJ...334..613S} classify NGC 3044 as having a low level of interaction, with companions within 10 D$_{25}$, but none that are clearly interacting.  This lack of nearby companions, however, does not rule out the possibility of past merger activity.  Alternatively, the disk could have been disturbed by an energetic event as suggested by extensive asymmetric extra-planar nonthermal emission observed by \citet{2013MNRAS.433.2958I} (see Figure 11 of that work).  Regardless, the morphology of NGC 3044 are inconsistent with those of an undisturbed disk.

\par
    DIG observations of NGC 3044 (among other galaxies) are presented by \citet{2000ApJ...536..645C}.  They note the asymmetric nature of the galaxy, as well as extra-planar DIG extending to roughly 1 kpc above the mid-plane.

\par
     \citet{1997ApJ...490..247L} previously observed and modeled NGC 3044 using the VLA in C configuration.   Their results will be discussed in $\S$~\ref{ngc3044obs}.  We will discuss our observations in detail in the next section.

\begin{centering}
\begin{deluxetable*}{lrr}
\tabletypesize{\scriptsize}
\tablecaption{NGC 3044 Parameters\label{n3044tbl-1}}
\tablewidth{0pt}
\tablehead
{
\colhead{Parameter} &
\colhead{Value}&
\colhead{Reference}\\
}
\startdata
\phd Distance (Mpc) &18.7\tablenotemark{a} &NASA Extragalactic Database\\ 
\phd Systemic velocity (km s$^{-1}$)&1260 &This work\\
\phd Inclination &85$\,^{\circ}$ &This work\\
\phd SFR (M$_{\odot}$ yr$^{-1}$) &1.34\tablenotemark{b} & Rossa \& Dettmar 2000\\ 
\phd Morphological Type &SBc & \citet{1988JRASC..82..305T} \\
\phd Kinematic Center $\alpha$ (J2000.0) &  09h 53m 40.1s &This work\\
\phd Kinematic Center $\delta$ (J2000.0) & 01d 34m 46.7s &This work\\
\phd D$_{25}$ (kpc) &27.0& \citet{1991trcb.book.....D}\\
\phd Total Atomic Gas Mass ($10^{9}M_{\odot}$) &3.5  \tablenotemark{c}&This work\\
\enddata
\tablenotetext{a}{Distance is the median value of distances found on the NED database, excluding those obtained using the Tully-Fisher relation in order to be consistent with HALOGAS.}
\tablenotetext{b}{Adjusted from 0.71 to account for different distances used.}
\tablenotetext{c}{Includes neutral He via a multiplying factor of 1.36.  The value obtained using single dish data is 5.1$\times10^{9}M_{\odot}$ \citep{2005ApJS..160..149S}.}
\end{deluxetable*}
\end{centering}

\subsection{NGC 4302}

\begin{figure}
\centering
\includegraphics[width=80mm]{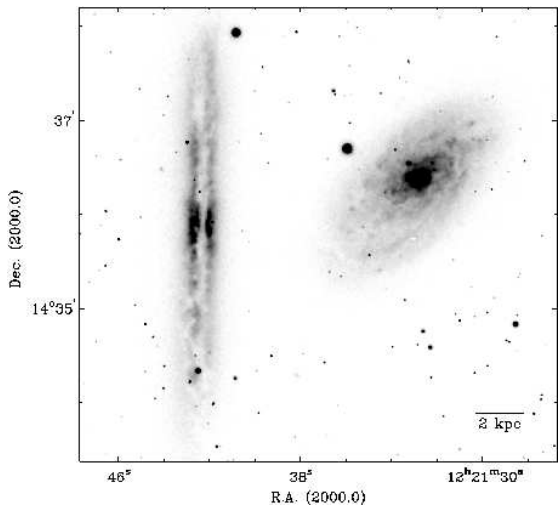}
\caption[NGC 3044 in Red Continuum]{\scriptsize\textit{A red image of NGC 4302 and NGC 4298 from \citet{1996ApJ...462..712R}.} \label{n4302_optical}}
\end{figure}

\par
    Basic properties of NGC 4302 are given in Table~\ref{n4302tbl-2}.  NGC 4302 is a nearby, edge-on spiral galaxy that is also a member of the Virgo Cluster.  It has a nearby companion (NGC 4298), and is likely undergoing ram pressure stripping due to the cluster environment \citep{2007ApJ...659L.115C}.  Both galaxies are shown in Figure~\ref{n4302_optical}.

\par
   \citet{2007ApJ...663..933H} observed and modeled the extra-planar DIG morphology and kinematics in NGC 4302.  In that work, kinematics in the western half of the galaxy were more easily constrained, although the kinematics of the east side did not significantly differ, thus a global DIG lag exists.  They find a lag of $-$39 km s$^{-1}$ kpc$^{-1}$ in the NW quadrant, and $-$26 km s$^{-1}$ kpc$^{-1}$, \textit{steepening} to $-$67 km s$^{-1}$ kpc$^{-1}$ near a radius of 3.8 kpc in the SW quadrant (assuming our adopted distance of 15.1 Mpc).  A radially steepening lag has not been seen in any other galaxy, making NGC 4302 an unusual case.  

\par
  {\sc H\,i} observations were carried out by \citet{2007ApJ...659L.115C} as part of a larger survey of the Virgo Cluster using the VLA in C configuration, although modeling was not done as part of that work.  We combine these observations with higher resolution B configuration observations in order to investigate the kinematics of the extra-planar {\sc H\,i}. 

\begin{centering}
\begin{deluxetable*}{lrr}
\tabletypesize{\scriptsize}
\tablecaption{NGC 4302 Parameters\label{n4302tbl-2}}
\tablewidth{0pt}
\tablehead
{
\colhead{Parameter} &
\colhead{Value}&
\colhead{Reference}\\
}
\startdata
\phd Distance (Mpc) &15.1\tablenotemark{a} &NASA Extragalactic Database\\ 
\phd Systemic velocity (km s$^{-1}$)&1150 &This work\\
\phd Inclination &90$\,^{\circ}$ &This work\\
\phd SFR (M$_{\odot}$ yr$^{-1}$) &0.9&\citet{2013AJ....145...62R} \\ 
\phd Morphological Type &Sc& \citet{1988JRASC..82..305T} \\
\phd Kinematic Center $\alpha$ (J2000.0) &  12h 21m 42.3s &This work\\
\phd Kinematic Center $\delta$ (J2000.0) & 14d 35m 50.1s &This work\\
\phd D$_{25}$ (kpc) &24.4& \citet{1991trcb.book.....D}\\
\phd Total Atomic Gas Mass ($10^{9}M_{\odot}$) &1.9  \tablenotemark{b}&This work\\
\enddata
\tablenotetext{a}{Distance is the median value of distances found on the NED database, excluding those obtained using the Tully-Fisher relation in order to be consistent with HALOGAS.}
\tablenotetext{b}{Includes neutral He via a multiplying factor of 1.36.  The value obtained using single dish data is 2.2$\times$10$^{9}$ M$_{\odot}$ \citep{2007AJ....133.2569G}.}
\end{deluxetable*}
\end{centering}

\section{Observations \& Data Reduction}

\subsection{NGC 3044}\label{ngc3044obs}

\par
   Observations of NGC 3044 were obtained with the VLA in both the B and C configurations (Table~\ref{3044tbl-2}).  The 8 kHz bandwidth was divided into 256 channels, yielding a velocity resolution of 31.25 kHz per channel (6.6 km s$^{-1}$ - increased to 6.7 km s$^{-1}$ during data reduction in order to properly combine all tracks).  

\par
    Data reduction was performed in {\tt CASA} using standard spectral line methods.  Self-calibration was performed after combining all tracks (with the exception of archival data) but before continuum subtraction.  This yielded a substantial reduction in sidelobes, as well as improved detection of faint, extended emission.  A variety of \textit{uv} weighting schemes were tried, in the end Briggs weighting with a somewhat high robust parameter of 2 was used for our final (12.3$\times$11.1" resolution) cube with an rms noise of 0.28 mJy bm$^{-1}$ in a single 6.7 km s$^{-1}$ channel.  Primary beam correction was applied to all cubes using the {\tt CASA} task {\tt IMMATH}.

\par
     Zeroth-moment maps are created by first using the {\tt GIPSY} task {\tt PYBLOT} to exclude excess regions of noise, and then using {\tt MOMENTS}.

\begin{deluxetable*}{lr}
\tabletypesize{\scriptsize}
\tablecaption{NGC 3044 Observational and Instrumental Parameters  \label{3044tbl-2}}
\tablewidth{0pt}
\tablehead
{
\colhead{Parameter} &
\colhead{Value}\\
}
\startdata
\phd Observation Dates$-$C Configuration &2010 Nov 21\\
\phd &2010 Nov 22\\
\phd &2010 Nov 23\\
\phd &2010 Nov 26\\
\phd &2010 Nov 30\\
\phd &2010 Dec 06\\
\phd B Configuration &2011 Mar 20\\
\phd &2011 Apr 07\\
\phd &2011 May 01-02\\
\phd &2011 May 04-05\\
\phd Total On-Source Time in C Configuration (hours)&13\\
\phd Total On-Source Time in B Configuration (hours)&11.5\\
\phd Pointing Center &09h 53m 40.3s\\
\phd &01d 34m 48.9s\\ 
\phd Number of channels &256\\
\phd Velocity Resolution &6.7 km s$^{-1}$\\

\phd Beam Size &12.27$\times$11.14"\\
\phd &1100$\times$1000 pc\\
\phd PA &$-$25.0$^{\circ}$\\

\phd RMS Noise $-$ 1 Channel (12.27$\times$11.14")&0.28 mJy bm$^{-1}$\\

\enddata
\end{deluxetable*}

\subsection{NGC 4302}

\par
   Observations were performed using the VLA in both B and C configurations, with the latter being archival, originally presented in \citet{2009AJ....138.1741C}.  Observational parameters are listed in Table~\ref{4302tbl-2}.  

\begin{deluxetable*}{lr}
\tabletypesize{\scriptsize}
\tablecaption{NGC 4302 Observational and Instrumental Parameters  \label{4302tbl-2}}
\tablewidth{0pt}
\tablehead
{
\colhead{Parameter} &
\colhead{Value}\\
}
\startdata
\phd Observation Dates $-$C Configuration &10-11 July 2005\\
\phd B Configuration &22 Apr 2009\\
\phd &24-25 Apr 2009\\
\phd Total On-Source Time in C Configuration (hours)&8\\
\phd Total On-Source Time in B Configuration (hours)&24\\
\phd Pointing Center &12h 21m 37.6s\\
\phd &14d 36m 7.0s\\ 
\phd Number of channels &63\\
\phd Velocity Resolution &20.1 km s$^{-1}$\\
\phd Beam Size (primary cube) &10.9$\times$9.1"\\
\phd &800$\times$670 pc\\
\phd PA &$-$10.1$^{\circ}$\\
\phd Beam Size (smoothed cube) &30$\times$30"\\
\phd &2200$\times$2200 pc\\
\phd RMS Noise $-$ 1 Channel (10.9$\times$9.1")&0.15 mJy bm$^{-1}$\\
\phd RMS Noise $-$ 1 Channel (30$\times$30")&0.048 mJy bm$^{-1}$\\

\enddata
\end{deluxetable*}

\par
   Data reduction was performed in {\tt AIPS} using standard spectral line methods.  Self calibration was performed after continuum subtraction on averaged channels in order to obtain a sufficient SNR, which yielded moderate improvement.  A variety of weighting schemes were used in the imaging process.  The final, full resolution cube was created using Briggs weighting with a robust parameter of 3, which was found to be the best balance between resolution and sensitivity after experimenting with the full range of robust parameters available in AIPS.  Our final cube used for all modeling has a spatial resolution of 10.9$\times$9.1" and an rms noise of 0.15 mJy bm$^{-1}$ in a single 20.1 km s$^{-1}$ channel.  To investigate the bridge between NGC 4302 and NGC 4298, we create a smoothed 30$\times$30" cube.

\par
   To investigate the origins of the extra-planar {\sc H\,i} in these two galaxies, we employ tilted ring modeling.  With tilted ring modeling (described in $\S$~\ref{modeling_strategy}), we can explore the morphology and kinematics both within the midplane, as well as in the extra-planar layers.

\section{Overview of Modeling Strategy}\label{modeling_strategy}
\par
   Using the Tilted Ring Fitting Code ({\tt TiRiFiC}, \citealt{2007A&A...468..731J}), models comprised of concentric tilted rings are created for each galaxy beginning with the simplest case: a layer fit by a single exponential (1C) in the vertical direction.  To avoid the addition of unnecessary free parameters, additional global features such as warps (W), flares, and second, thicker components (2C) are added only when necessary to match the data and \textit{the simplest model that fits the data is chosen}.  Lags (L) are only considered if the kinematics as a function of height cannot be fit by such models and indicate that a lag might be present.  Finally, in some cases, {\tt TiRiFiC} allows for the fitting of localized features (F) and asymmetries.  Additionally, we are able to add radial motions (R) to the models.

\par
    Due to asymmetries and localized features, the models must be compared to the data by eye. Comparisons are made between model and data minor and major axis position-velocity diagrams (bv and lv diagrams, respectively), as well as channel maps.  While only representative diagrams are included here, the entire cube is evaluated for each model. Uncertainties on final values of each parameter are estimated by varying it with all other parameters fixed until the model clearly disagrees with the data.  For a more in-depth assessment of the uncertainties involved in tilted-ring modeling and a further explanation of this technique see ~\citet{2012ApJ...760...37Z} and ~\citet{2013MNRAS.434.2069K}.

\section{NGC 3044}
\subsection{The Data}

\par
   NGC 3044 was previously observed in {\sc H\,i} by \citet{1997ApJ...490..247L} in C configuration only. To these data we add higher resolution B configuration data as well as additional C configuration data to increase sensitivity. Most of the noteworthy features we see upon inspection are also identified in their paper - it is later in the modeling stages where our data provide additional constraints.  Thus, we only briefly mention the most prominent features and refer the reader to \citet{1997ApJ...490..247L} for greater detail when appropriate, especially in the case of {\sc H\,i} shells, since that was a prime focus of their work. 

\begin{figure}
\centering
\includegraphics[width=80mm]{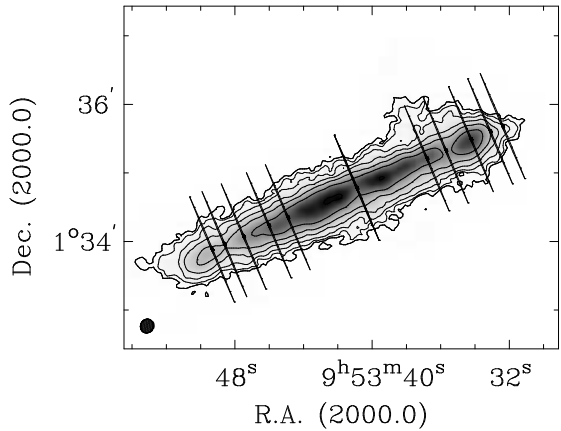}
\caption{\scriptsize\textit{A zeroth-moment map of NGC 3044.  Contour levels begin at 1.6$\times$10$^{20}$ cm$^{-2}$ and increase by factors of 2.  Lines represent the slice locations and minor-axis widths of bv diagrams in Figure~\ref{n3044bvapproachingpaper2}.  The beam is shown in the lower left-hand corner.  The northwest half is the approaching half.  Note the apparent thickness of the {\sc H\,i} layer, including extra-planar features.  Also note the asymmetry in {\sc H\,i} about the dynamical center (the middle slice is taken at the kinematic center).} \label{totalHIdata3044}}
\end{figure}

\par
\begin{figure}
\centering
\includegraphics[width=80mm]{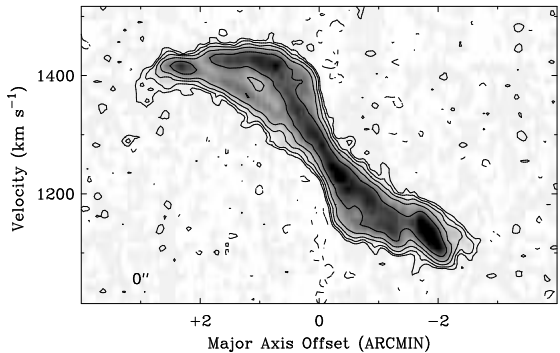}
\caption[A Major Axis Position-velocity Diagram of NGC 3044]{\scriptsize\textit{An lv diagram along the major axis of NGC 3044.  Note the flatness of the terminal side of the receding half, and the slant on the approaching half that brings velocities at small to moderate radii closer to the systemic velocity relative to the receding side at the same offsets. Also note the slight difference in spacing between the contours on the terminal sides of each half, indicated that the approaching half has a higher velocity dispersion.  The difference in shape of the two terminal sides as well as the clear difference in velocity dispersion indicate that the asymmetries cannot be fixed by shifting the systemic velocity, but rather can be attributed to kinematic differences between the two halves, possibly driven by a disturbance.  Contours begin at 2$\sigma$ (0.56 mJy bm$^{-1}$, or 3.0$\times$10$^{19}$ cm$^{-2}$) and increase by factors of two.  Negative 2$\sigma$ contours are shown as dashed lines} \label{3044lvmajor2}}
\end{figure}

   Immediately evident from the zeroth-moment map (Figure~\ref{totalHIdata3044}) is the apparent thickness of the {\sc H\,i} disk, as well as the substantial lop-sidedness.  This lop-sidedness is also seen in the lv diagram displayed in Figure~\ref{3044lvmajor2}, which also shows the differing kinematics of the two sides.  Also noteworthy (Figure~\ref{totalHIdata3044}) are several features also observed by \citet{1997ApJ...490..247L}, the most striking being the extra-planar extension in the NW quadrant of the galaxy, which corresponds to ``F10" in their work.  

\par
\begin{figure}
\centering
\includegraphics[width=80mm]{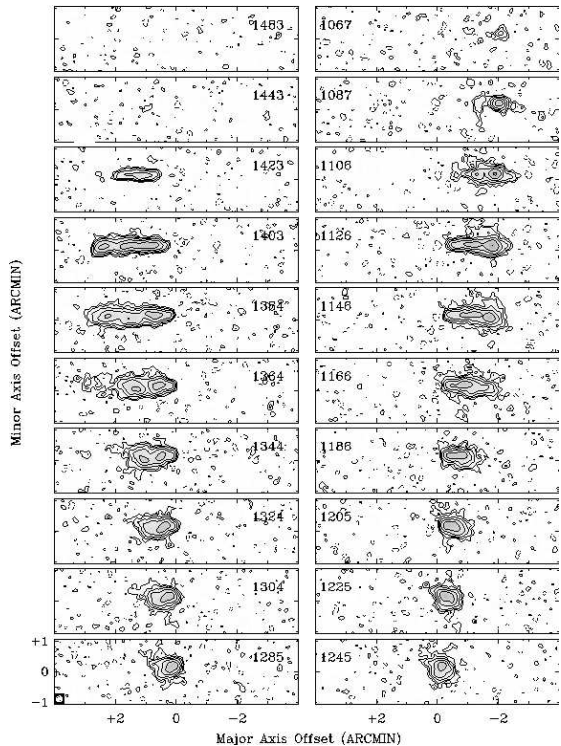}
\caption[Channel Maps of NGC 3044]{\scriptsize\textit{Channel maps of NGC 3044 in a rotated frame with the origin at the kinematic center show the substantial asymmetries within the galaxy. Note the extra-planar emission at positive z in panels corresponding to 1106-1205 km s$^{-1}$.  Velocities are given for each panel.  Central channels are shown in Figure~\ref{3044channelmapcenter2}. Contours are presented as in Figure~\ref{3044lvmajor2}.} \label{n3044channelmapsymmetricpaper2}}
\end{figure}

   Upon initial inspection of Figure~\ref{totalHIdata3044}, it can be seen that the kinematic and large scale morphological (determined by the {\sc H\,i} extent along the major axis) centers differ by approximately 0.4' (2.2 kpc) (the central black line defines a slice through the cube taken at the kinematic center we determine later).  There are indications in the outer parts of the disk that there may be a warp component across the line of sight, although modeling is required to say for sure.  The overall clumpiness of the data indicate probable contributions from spiral structure.  These properties may also be seen in the channel maps of Figure~\ref{n3044channelmapsymmetricpaper2}.

\subsection{The Modeling of NGC 3044}\label{3044modeling}

\par
\begin{figure}
\centering
\includegraphics[width=80mm]{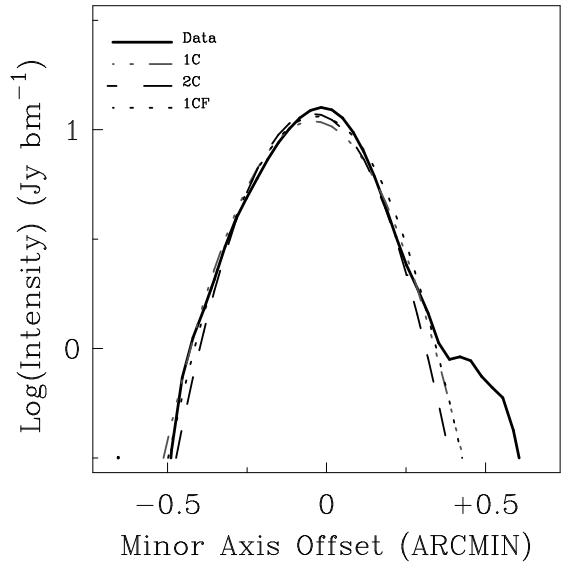}
\caption{\scriptsize\textit{Vertical profiles of NGC 3044 (data), one-component (1C), two-component (2C), and (1CF; where all models beginning with ``1CF" are represented by this profile) models summed over a region of $\pm$40" (3.6 kpc) from the kinematic center.  The 1C and 1CF curves lie nearly on top of each other in the wings, but then diverge near the peak. Each model is iteratively optimized to fit pv diagrams, channel maps, etc. in addition to the vertical profiles, resulting in somewhat of a compromise in the fits seen in any one figure.  Furthermore, since it is clear at an early stage that a 2C model is not warranted, it has undergone fewer iterations than the 1C models, and thus the fit shown for the 2C model is not fully-optimized, but still illustrates that it does not provide a sufficient improvement to justify the additional free parameters.  Note the extension on the positive side that corresponds to the high-z emission to the North in Figure~\ref{n3044channelmapsymmetricpaper2} that we do not attempt to model.} \label{n3044vprofsumlog2}}
\end{figure}

   The position angle of the main disk of NGC 3044 was determined by eye to be 113$^{\circ}$ (Figure~\ref{totalHIdata3044}) and the initial inclination was assumed to be 90$^{\circ}$ (refined to a final value of  85$^{\circ}$$^{+1}_{-0.5}$).  Initial estimates for the rotation curve were based on work presented in \citet{1997ApJ...490..247L}.  These quantities were later adjusted by eye starting with examination of lv-diagrams, followed by channel maps.  The surface brightness distribution was originally obtained using the {\tt GIPSY} task {\tt RADIAL}, which determines a radial surface density distribution from a major axis emission profile derived from a zeroth-moment map of an edge-on or nearly edge-on galaxy.  The {\sc H\,i} scale height was determined by examining the vertical profile and zeroth-moment maps (Figures~\ref{n3044vprofsumlog2} and~\ref{totalHIdata3044}).  The initial velocity dispersion (estimated by viewing the spacing of contours on the terminal sides of bv diagrams) is 20 km s$^{-1}$ throughout the disk of the approaching half, and 20 km s$^{-1}$ at the center of the receding half, but decreasing to 10 km s$^{-1}$ by a radius of 1.3'.  The higher velocity dispersion in the approaching half is related to the somewhat large spread of velocities on the terminal side most readily inferred from lv diagrams (Figure~\ref{3044lvmajor2}). It is clear at an early stage that there exists a warp component perpendicular to the line of sight which we include in all of our models.  Finer adjustments to these parameters were made iteratively throughout the modeling process.

\par
\begin{figure}
\centering
\includegraphics[width=80mm]{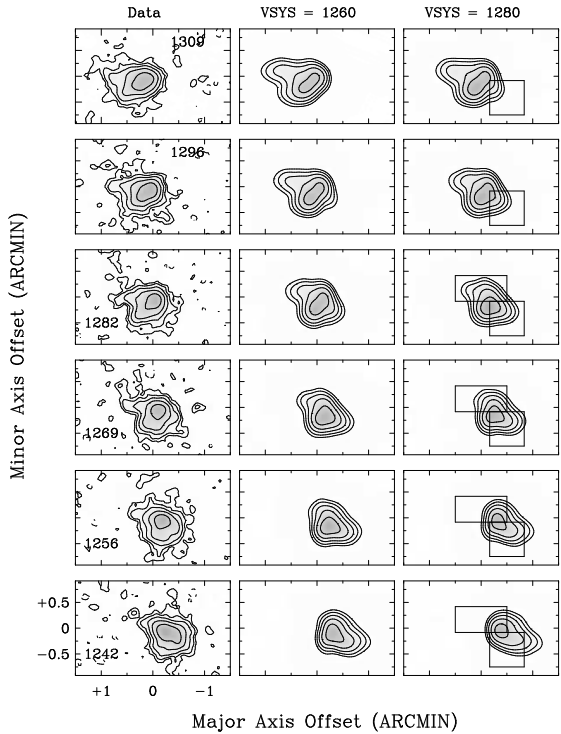}
\caption[Central Channel Maps of NGC 3044]{\scriptsize\textit{Central channel maps of NGC 3044 showing the data, the final model in the center (VSYS = 1260 km s$^{-1}$), and a model with a systemic velocity of 1280 km s$^{-1}$.  Velocities in km s$^{-1}$ are given in each panel. The major axis off-set is with respect to the kinematic center chosen for our models. Neither model is an excellent fit as the central regions of NGC 3044 are abnormally complicated.  However, the 1260 km s$^{-1}$ is a better fit, although the differences are subtle.  Note how the shape of all data channels at velocities of 1282 km s$^{-1}$ and higher appear as though they should be a part of the receding half, indicating that the systemic velocity must be lower than 1282 km s$^{-1}$.  This velocity range was chosen as it provides the clearest examples showing that a systemic velocity of 1280 km s$^{-1}$ (or higher) would indeed be a poor fit to the data.  Regions with noticeably poor fits in the 1280 km s$^{-1}$ model are marked with boxes.} \label{3044channelmapcenter2}}
\end{figure}

   Due to asymmetries we model the two halves separately.  The systemic velocity was difficult to determine, in part because the kinematic and morphological centers are in different locations, which \citet{1997ApJ...490..247L} also noted.  Furthermore, the rotation curves differ greatly in their respective shapes as may be inferred from Figure~\ref{3044lvmajor2}.  However, with {\tt TiRiFiC}, we now have considerably more flexibility to model complicated kinematics and morphology.  In our first approach, we set the systemic velocity to 1280 km s$^{-1}$, a value for which the rotation curves in each half are within reasonable agreement (within 5-10 km s$^{-1}$ scatter of each other).  Our reasoning for this is that, for an undisturbed galaxy, it is more physically plausible that the two rotation curves are comparable to each other.  However, this approach fails to fit shapes of the central channels (Figure~\ref{3044channelmapcenter2}), and would require additional components (such as arcs or a bar, which were also attempted with varying degrees of success, a bar being the worst fit) to match the kinematics near the center.  Thus, to fit central channels, the systemic velocity is decreased to 1260 km s$^{-1}$ (Figure~\ref{3044channelmapcenter2}; even this choice still leads to some mismatches with the data, but was found to be the best value), which in turn causes a large discrepancy in the rotation curves of the two halves.  The same kinematic center is used in each case. Thus, this appears to be a true difference in the kinematics. Fortunately, a shift in the systemic velocity (provided the rotation curves are properly adjusted as well) primarily affects the central regions, so this has a minimal effect on final models.  A different systemic velocity could be used for outer rings, which would bring the rotation curves into better agreement, but for simplicity, we choose to allow only a single systemic velocity for all radii in our models.  More importantly, because we do not rely on central regions for measuring lags, such a shift in systemic velocity has no impact on lag values determined through modeling.

\subsubsection{Individual Models}

\par
   We now address individual models created in our analysis and their defining characteristics.

\par
     The best fit achieved with a one-component model (all parameters optimized) uses an exponential scale height of 7" (635 pc).  As can be seen in Figure~\ref{n3044vprofsumlog2}, if the symmetric extended emission in the wings is matched, the fit to the peak is somewhat compromised. (This extra-planar emission is not the large asymmetric extension at positive offsets, which we discuss later in this section.)  In an attempt to improve the fit, a two-component model with the thin disk having a scale height of 4" (360 pc), and the thick disk having a scale height of 8" (730 pc) is tested and found to be a comparable fit to the data, but yields no improvement.   Other combinations of scale heights were also investigated, but none of these could improve upon the fit.  Also in the bv diagrams, there was little improvement through the addition of a second component.  Thus, there is no compelling reason for a second component, especially given the increase in the number of free parameters, so it is omitted from further analysis.

\begin{figure}
\centering
\includegraphics[width=80mm]{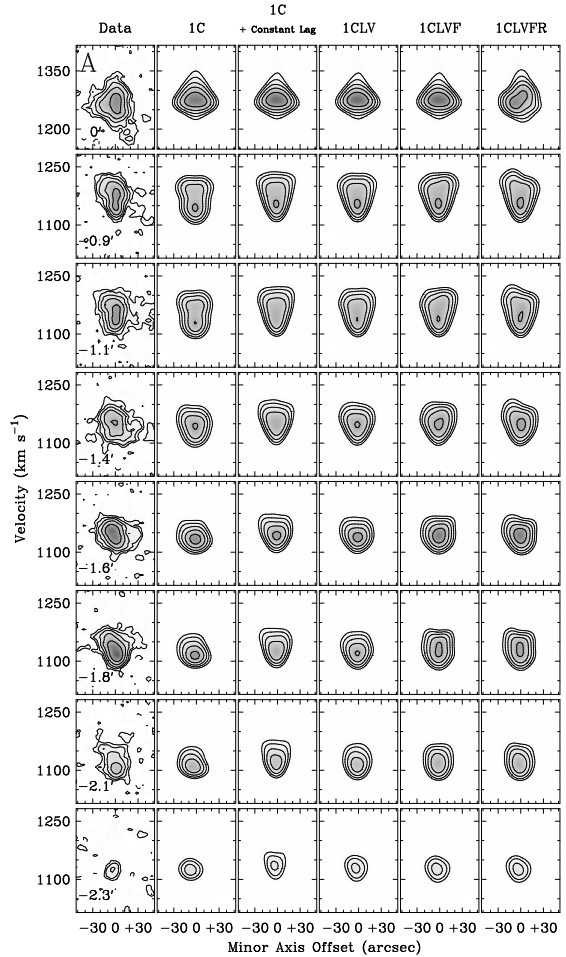}
\caption[Minor Axis Position-velocity Diagrams of NGC 3044 (Data and Models)]{\scriptsize\textit{bv diagrams of the approaching (A) and receding half (B) of NGC 3044. Note the improvement of the fit due to the radial shallowing of the lag.   Adding arcs to the model allows for an improved fit to the higher contours, but has no noticeable effect on the lag (column 4).  Finally, as may be seen in the final column, the slants on the systemic sides are brought into closer agreement with the data through the addition of radial motions.  Contours are presented as in Figure~\ref{3044lvmajor2}.  Slice locations are shown by the black lines in Figure~\ref{totalHIdata3044} and positive offsets correspond to the northern side.}\label{n3044bvapproachingpaper2}}
\end{figure}

\begin{figure}
\centering
\includegraphics[width=80mm]{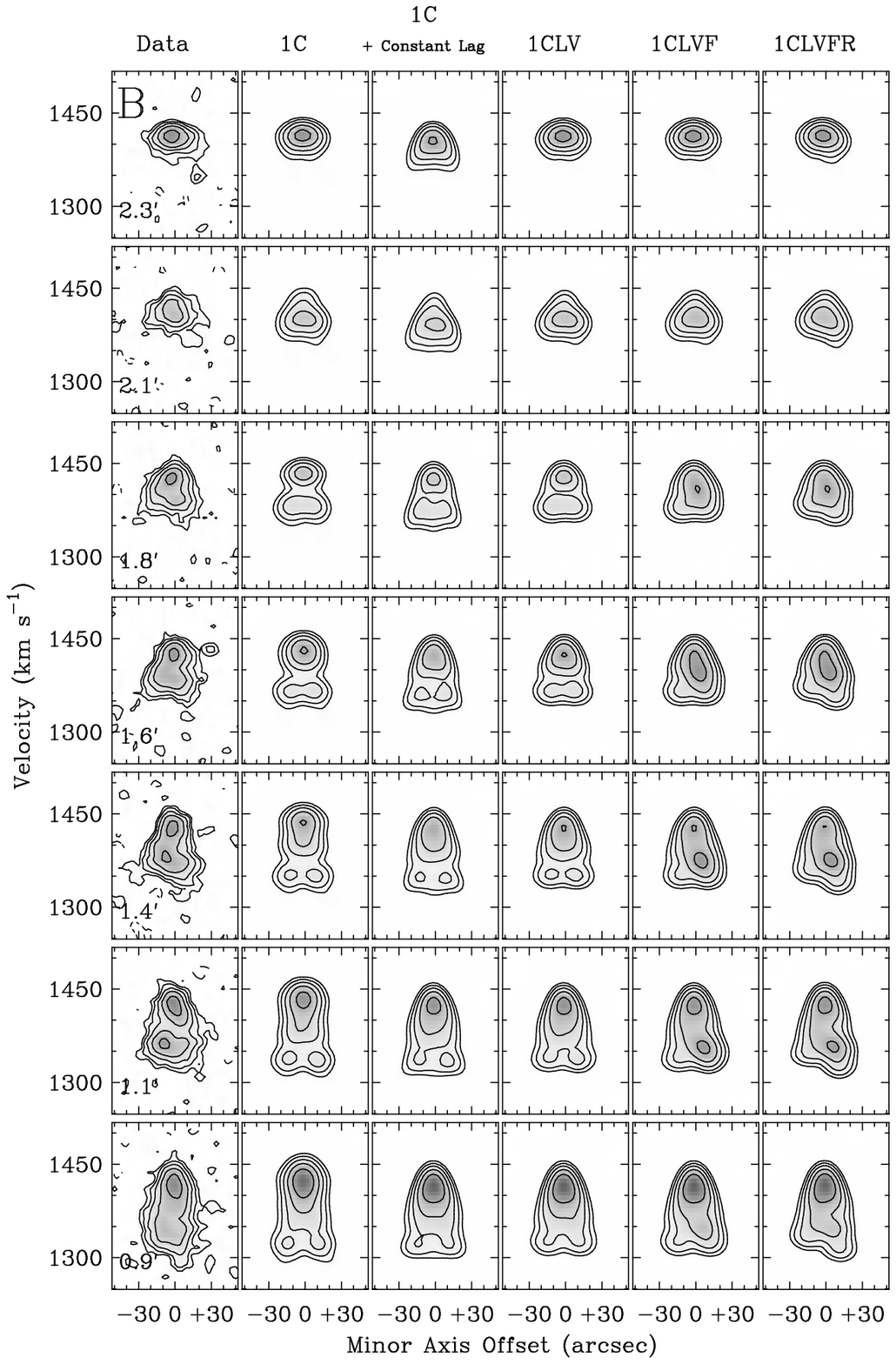}
\end{figure}

\begin{figure*}
\centering
\includegraphics[angle = 270, width = 150mm]{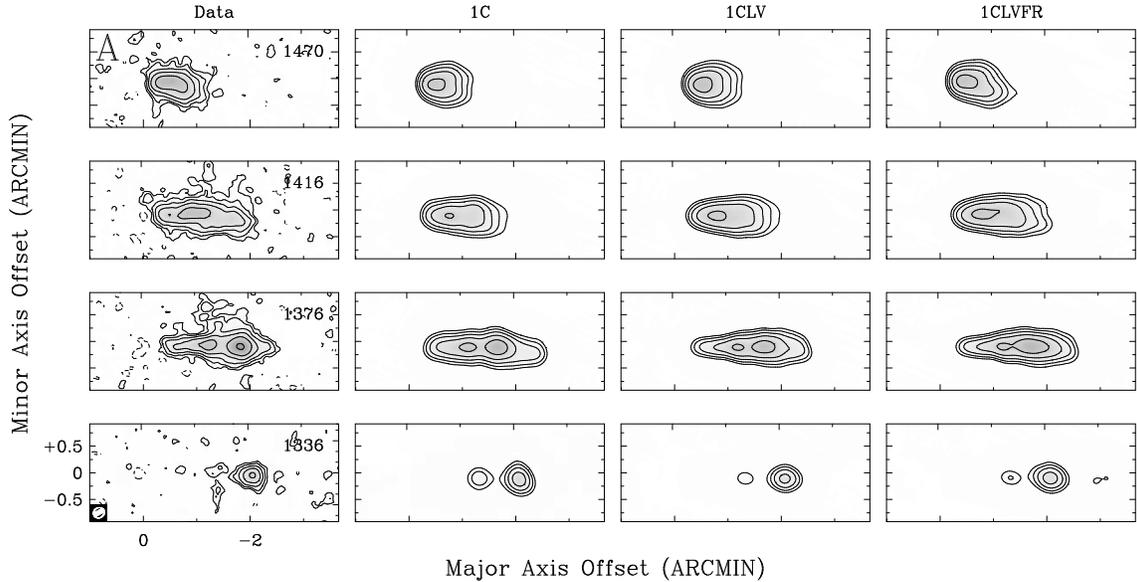}
\caption[Channel Maps of NGC 3044 and Models]{\scriptsize\textit{Representative channel maps of the approaching (A) and receding (B) halves of NGC 3044.  Velocities in km s$^{-1}$ are given in each panel.  The models presented here are the 1C, 1C with an optimal lag, and then our final 1CLFR model with arcs and radial motions.  Note the narrowing of the channel maps toward the center due to the lag.  The addition of radial motions replicates some of the slant seen in the data.  Also note the previously mentioned extra-planar emission at 1376 km s$^{-1}$ and 1416 km s$^{-1}$, which we do not model.  Contours are presented as in Figure~\ref{3044lvmajor2}.} \label{3044channelmapsapproaching2}}
\end{figure*}

\begin{figure*}
\centering
\includegraphics[angle = 270, width = 160mm]{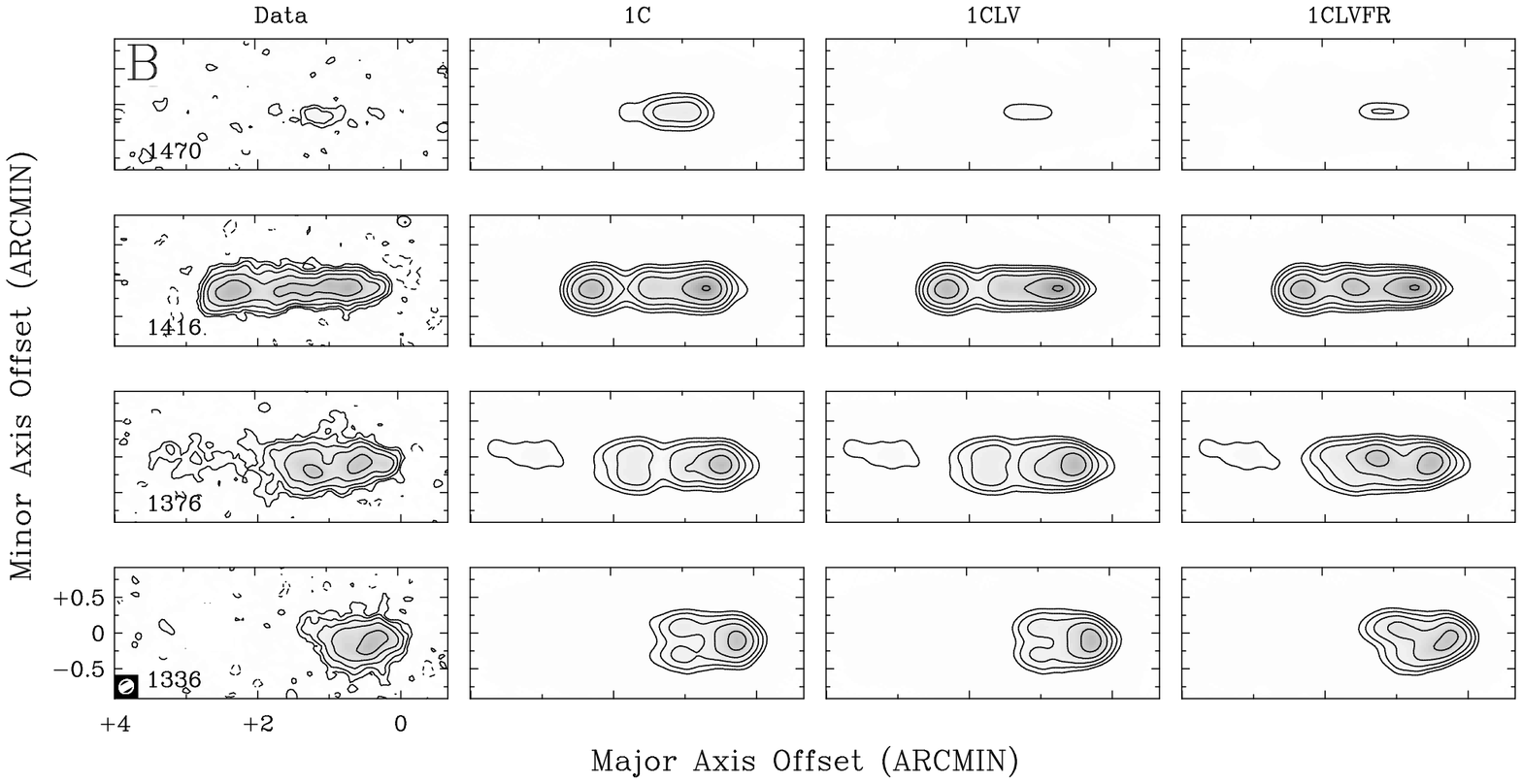}
\end{figure*}

\begin{figure}
\centering
\includegraphics[width=80mm]{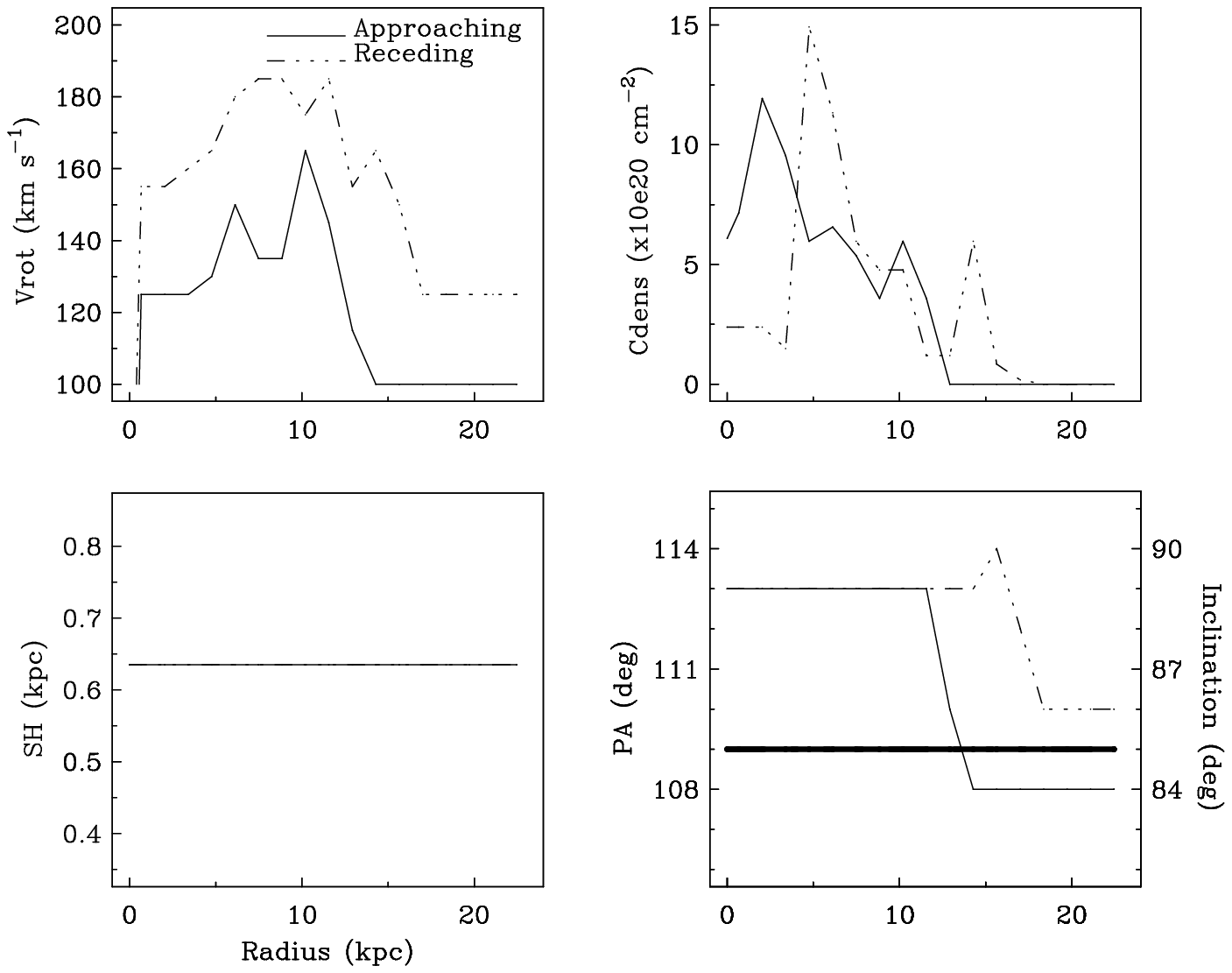}
\caption[Optimal Model Parameters for NGC 3044]{\scriptsize\textit{The parameters consistently used for all models of NGC 3044 presented here:  rotational velocities, column densities, scale height, inclination and position angle.  Note the discrepancies in the rotation curves of the two halves.  The bold line in the bottom-right panel represents the inclination in both halves.} \label{n3044.multiplot}}
\end{figure}

\par
   For the sake of completeness, we remark on the possibility of a warp component along the line of sight (or a decrease in overall inclination), which could also increase the apparent width of the disk, and even mimic a lag.  Fortunately, line of sight warp components show signature midplane indentations directed toward terminal velocities on the systemic sides of bv-diagrams (resulting in a subtle ``V" shape on the systemic side), as well as analogous features in the outer regions of channel maps, that are not seen in these data.  This effect can already be seen to some degree in the one component models in panels corresponding to an offset of $+$0.9'-1.4' in Figure~\ref{n3044bvapproachingpaper2}, and decreasing the inclination worsens it.

\par
   Figure~\ref{3044channelmapsapproaching2} displays channel maps of selected models. Parameters for the final models are given in Figure~\ref{n3044.multiplot}.

\subsubsection{Adding a Lag to the Models}

\par
\begin{figure}
\centering
\includegraphics[width=80mm]{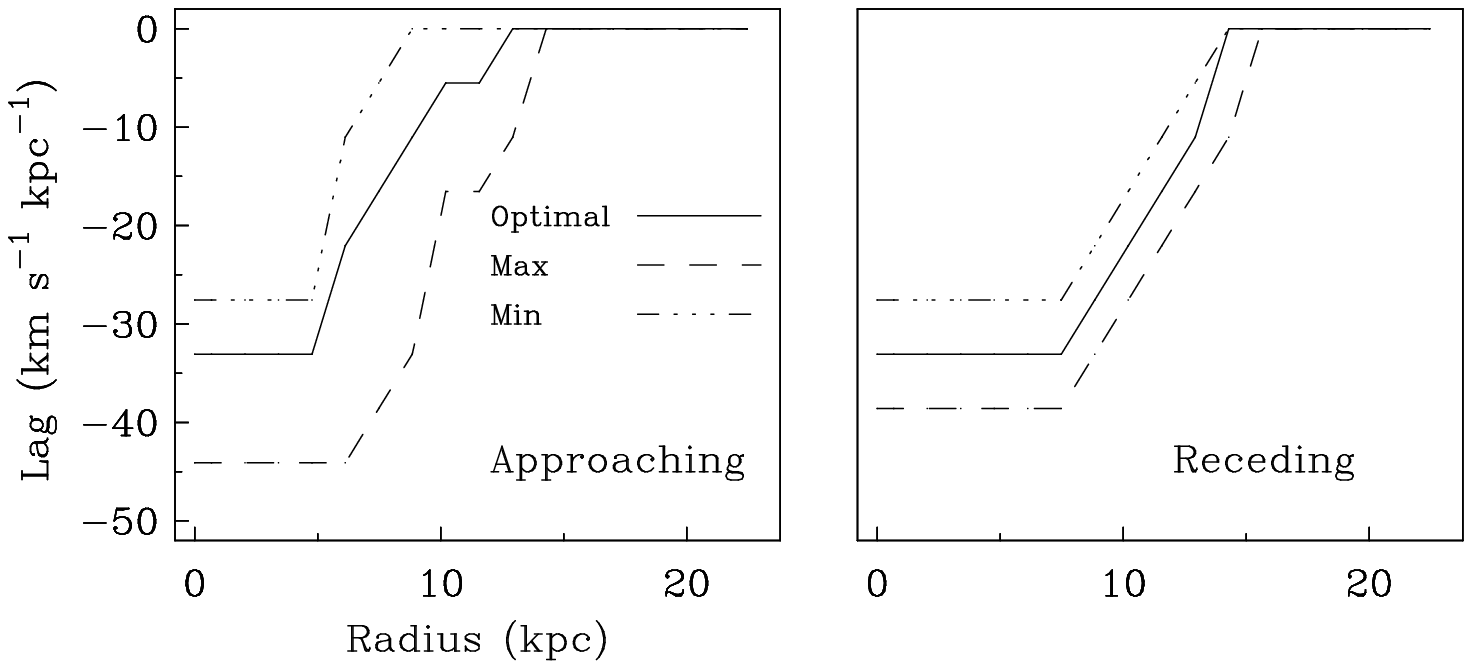}
\caption[The Lag Distribution in NGC 3044]{\scriptsize\textit{The optimal lag distributions along with the range of uncertainties for lag values in NGC 3044.  The uncertainties are quite high near the center, where projection effects, as well as the complicated nature of the central regions of NGC 3044 render precise modeling difficult.  The lag distributions in each half are quite similar, and go to zero near $R_{25}$ (2.5' or 13.6 kpc).} \label{3044parameterslag}}
\end{figure}
\begin{figure}
\centering
\includegraphics[width=80mm]{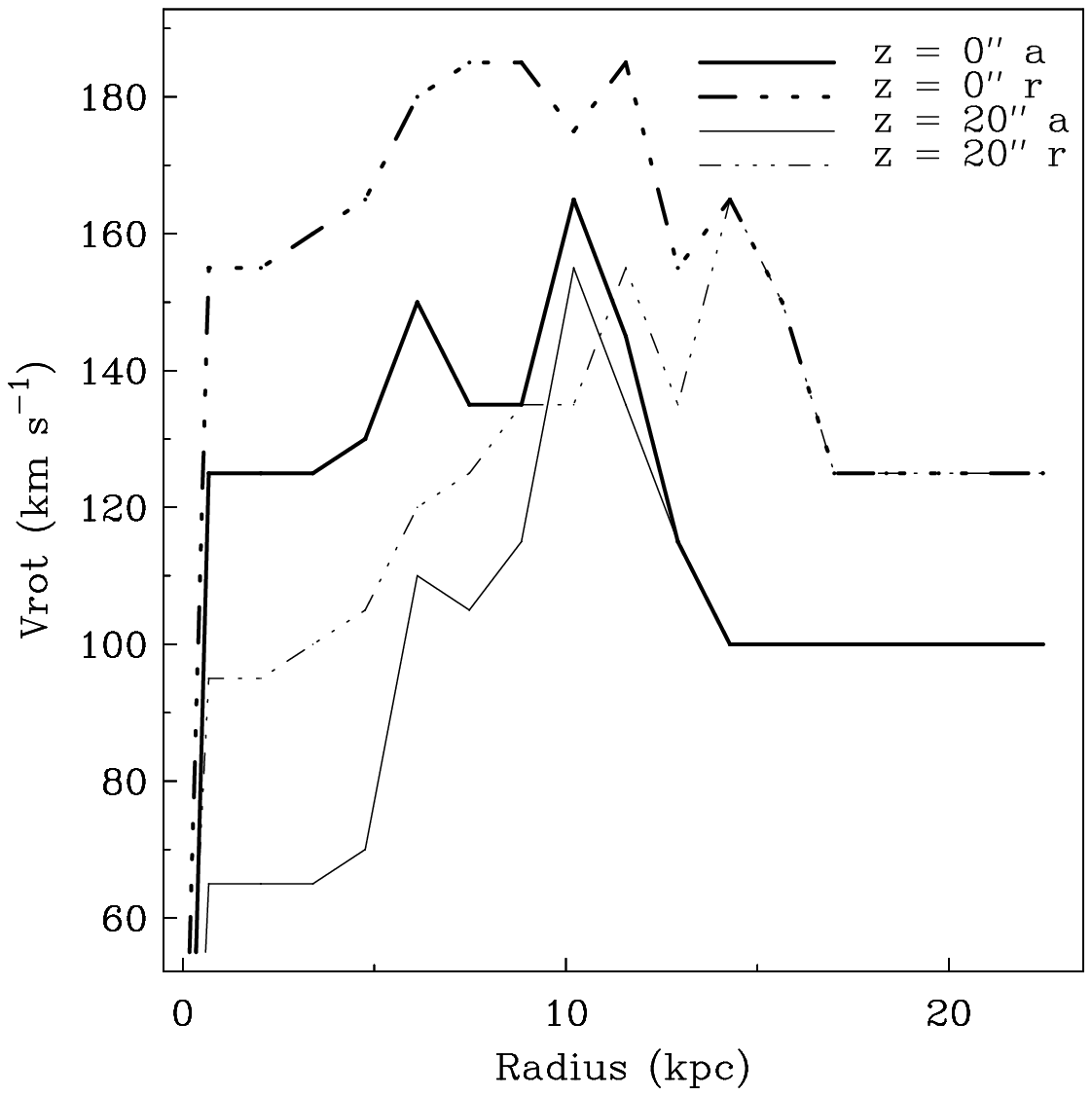}
\caption[The Rotation Curves at Varying Heights in NGC 3044]{\scriptsize\textit{The rotation curves at the midplane, as well as at z=20" (1.8 kpc) in NGC 3044.  The approaching half is indicated by ``a" and the receding half is indicated by ``r."}\label{3044parameterrotationcurves}}
\end{figure}

   Figure~\ref{n3044bvapproachingpaper2} shows a single component model (1C), single component with a constant lag (1C + constant lag), single component with a radially varying lag (1CLV), that model with additional features described in $\S$~\ref{distinct} (1CLVF), and a final model with radial motions added (1CLVFR). By examining the final three columns of Figure~\ref{n3044bvapproachingpaper2}, especially at $\pm$0.9' and $\pm$1.6', one can see that by adding an optimally determined lag of $-$33 $^{+6}_{-11}$ km s$^{-1}$ kpc$^{-1}$ in the approaching half, and $-$33 $\pm{6}$ km s$^{-1}$ kpc$^{-1}$ in the receding, the once rather boxy (i.e.\ the distribution of flux is concentrated near both the terminal and systemic sides even at large minor axis off-sets) model bv-diagrams are now more of a smooth ``V"-shape, in significantly better agreement with the data.  In Figure~\ref{3044channelmapsapproaching2} it may be seen that the models with no lag are too thick along the minor axis near the central regions compared to the data and adding a lag has a thinning effect.  While this global lag does improve the fit to the data, it is clear from Figure~\ref{n3044bvapproachingpaper2} that the lag is too steep at large major axis offsets.  By allowing the lag to shallow with radius, we see an improvement to the fit in the outer regions of the galaxy (column 4 of Figure~\ref{n3044bvapproachingpaper2}, column 3 of Figure~\ref{3044channelmapsapproaching2}).  As may be seen by examining the more rounded curvature on the terminal side of the bv diagrams at large radii, the model with a radially varying lag is a reasonable representation of the data.  We show the radial variations of the lag and its uncertainties (those quoted above are only near the center, with the uncertainties decreasing with radius) in Figure~\ref{3044parameterslag} and rotation curves in the disk and at z=20" in Figure~\ref{3044parameterrotationcurves}.

\subsubsection{Modeling Radial Motions and Distinct Features}\label{distinct}

\par
   In our final models, we add radial motions, as well as three distinct arcs to improve the fit.  We find that these features do not noticeably affect the lag values. 

\par
   There exist slants along the systemic side in the bv-diagrams (primarily those within 1.8').  These are modeled with radial motions of 15 km s$^{-1}$ for the entire approaching half, and 10 km s$^{-1}$ outside a radius of 2' (11.6 kpc) on the receding half.  In smaller radii on the receding half, the addition of radial motions having a positive sign are detrimental, and we see a very slight improvement by instead adding radial motions of $-$15 km s$^{-1}$ to radii smaller than 100" (9.1 kpc).  Improvement from these radial motions (largely due to those in outer radii) may also be seen by the analogous slants introduced in the final column of Figure~\ref{3044channelmapsapproaching2}.  It should be made clear that due to projection effects, the true sign of these motions cannot be determined (i.e.\ inflow vs. outflow) in absolute certainty.  However, the dusty extension below the plane of the disk on the approaching half in Figure~\ref{n3044redcont} indicates that the SW side (positive in bv diagrams) is the near side, and we quote the radial motions under this assumption.

\par
  The clumpiness of the data along the mid-plane indicates that, as to be expected, spiral structure is likely present.  Additionally, during the modeling process, it became clear that some of these features cannot be modeled by entire rings, and instead require the addition of arcs extending over only parts of the disk to properly fit the data.  We add such arcs sparingly to avoid unnecessarily increasing the complexity of the models.  Thus, not every clump is modeled, but can still be reasonably assumed to be due to spiral structure, or possibly merger activity.  Note that, due to projection effects, we do not claim these are unique representations of these features, but rather approximations that allow us to more closely match the data, and to show that they have little effect on our derived lag values.  We briefly describe the arcs included in our models below.

\par
\begin{figure}
\centering
\includegraphics[width=80mm]{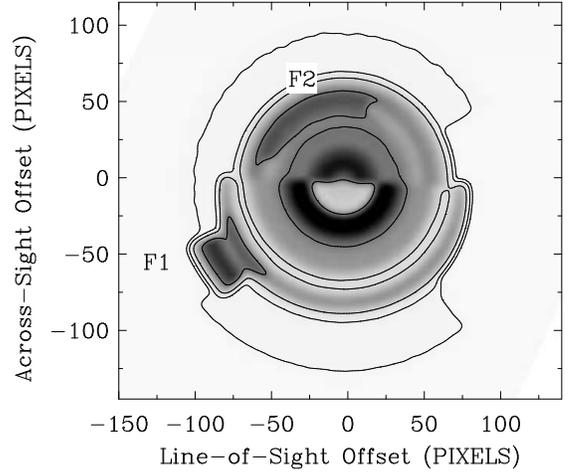}
\caption{\scriptsize\textit{A face-on view of the final model for NGC 3044 including all distinct features.  Based on our assumptions, the observer is oriented to the right of the figure, although some ambiguity exists due to projection effects, so the left cannot be ruled out.  ``F1" denotes the first arc described in this section, while ``F2" denotes the second.  Note that the third feature is too faint to be seen in this map.  Pixels are equivalent to 2" (180 pc).} \label{totalHIdataface3044}}
\end{figure}

  In Figure~\ref{n3044bvapproachingpaper2}, one may note that there is more flux present near the systemic side in the central panels (0.9'-1.8') in the data compared to the models, especially at positive minor axis off-sets.  This property cannot be replicated by increasing the surface brightness of entire rings, as doing so would over-saturate the systemic sides at large radii (not shown).  Thus, one arc, centered at an azimuth in the disk of 120$^{^\circ}$ (the approaching half is centered at 0$^{^\circ}$ for reference), having an initial azimuthal width of 40$^{^\circ}$ while decreasing radially to 20$^{^\circ}$, is added between 2.6' and 3.4' (14.5 and 18.1 kpc) radius.  The rotational velocity of this arc begins at 200 km s$^{-1}$ for smaller radii, and increases radially to 245 km s$^{-1}$.  The inclination and position angles are 85$^{^\circ}$ and 118$^{^\circ}$, respectively.  The surface brightness distribution assigned approximately doubles the total surface brightness in this region.  The scale height, as well as the two arcs described in the text, is 7" (630 pc).  A face-on view of the final model including the features described here may be seen in Figure~\ref{totalHIdataface3044}.

\par
   By examining Figure~\ref{n3044bvapproachingpaper2}, it can be seen that in the data compared to the models, there is more emission in the higher contours of panels corresponding to 1.6' and 1.8'.  To improve the fit, a second arc extends from 1.6' to 1.8' (8.8-10.2 kpc) radius, and is centered at an azimuth of 30$^{^\circ}$ with an azimuthal width of 90$^{^\circ}$.  Radial motions of 15 km s$^{-1}$ are added, as well as a lag of $-$22 km s$^{-1}$ kpc$^{-1}$.  Again, this feature approximately doubles the surface brightness of this region.  The rotational velocity begins at 115 km s$^{-1}$ for smaller radii and increases radially to 125 km s$^{-1}$.  The inclination and position angles of this arc are 85$^{^\circ}$ and 113$^{^\circ}$.

\par
   Finally, in panels from $-$0.9'to $-$2.1' of Figure~\ref{n3044bvapproachingpaper2}, the excess of emission on the systemic side in the upper, left-hand corner of the data does not slant in a smooth fashion, but retains some rounding on the right-hand side.  To duplicate this curvature in the models, a final arc centered at an azimuth of 30$^{^\circ}$, with a width of 180$^{^\circ}$ is also included.  The radial extent is from 2.4' to 3.4' (12.9-18.1 kpc).   The rotational velocity of this arc is 165 km s$^{-1}$, while the inclination and position angles are 85$^{^\circ}$ and 113$^{^\circ}$. This arc is fainter than the previous two, only adding approximately 20$\%$ more flux to the total for this region.  The effects of this arc can be most plainly seen in column 5 of Figure~\ref{n3044bvapproachingpaper2}, but the best overall fit can be seen in column 6 when it is combined with radial motions.  Improvements to the channel maps due to each of the above features can be seen in Figure~\ref{3044channelmapsapproaching2}.  Note the additional clumpiness and increased agreement in the density distribution compared to the data.  Finally, note the improved fit to the vertical profile (Figure~\ref{n3044vprofsumlog2}) that is due to the additional features.

\subsubsection{The Final Models and Their Uncertainties}
\par
\begin{figure}
\centering
\includegraphics[width = 80mm]{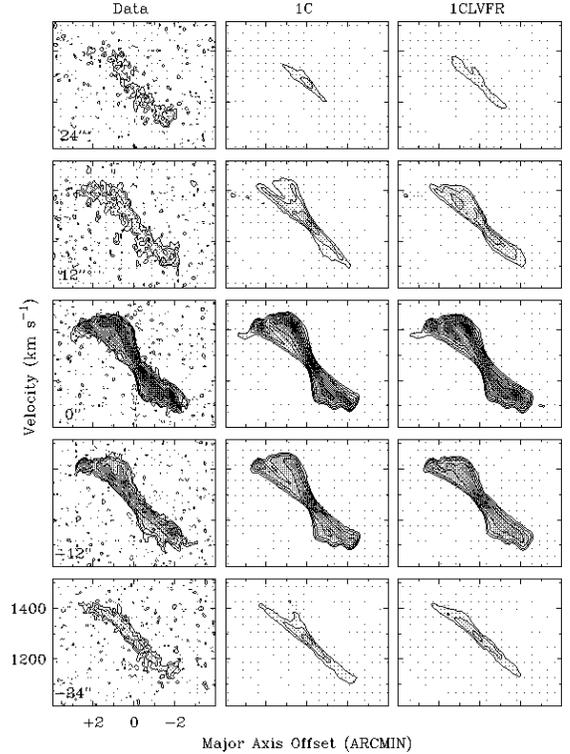}
\caption{\scriptsize\textit{Major axis position-velocity (lv) diagrams showing the data, one component (1C), and optimal models for NGC 3044. Offsets from the midplane are given in each panel and positive offsets correspond to the northern side. Although radial motions and additional features are included in the final model, almost all of the improvements seen here are due to the addition of a radially varying lag.  Contours are presented as in Figure~\ref{3044lvmajor2}.} \label{3044lvpaper2}}
\end{figure}

    The final models include a single component with a scale height of 7 $\pm$0.5" (635 $\pm$45 pc), a systemic velocity of 1260 $\pm$5 km s$^{-1}$, a radially shallowing lag with a maximum value of $-$33 $^{+6}_{-11}$ km s$^{-1}$ kpc$^{-1}$ in the approaching half, and $-$33 $\pm{6}$ km s$^{-1}$ kpc$^{-1}$ in the receding, three additional arcs, and radial motions described in the previous section.  Lv diagrams showing these models can be seen in Figure~\ref{3044lvpaper2} along with a comparison to a 1C model and the data.  The uncertainties in the rotation curves are approximately 5 km s$^{-1}$ for individual rings.  While we have attempted to quantify as many uncertainties as possible, the somewhat subjective nature of tilted-ring modeling makes absolute error estimates impossible.  However, we examine the residuals  and the sums of their squares of the 1C, 1CL, 1CLFR, and our final model, and find a substantial decrease between early and final models.  The percent decrease in the sum of the square of the residuals is 6.4\% between the 1C and 1CL models, 15.2\% between the 1C and 1CLFR models, and 37.7\% between the 1C and final models.  This steady decrease demonstrates improvement between each successive model.

\section{NGC 4302}
\subsection{The Data}
\par
\begin{figure}
\centering
\includegraphics[width=80mm]{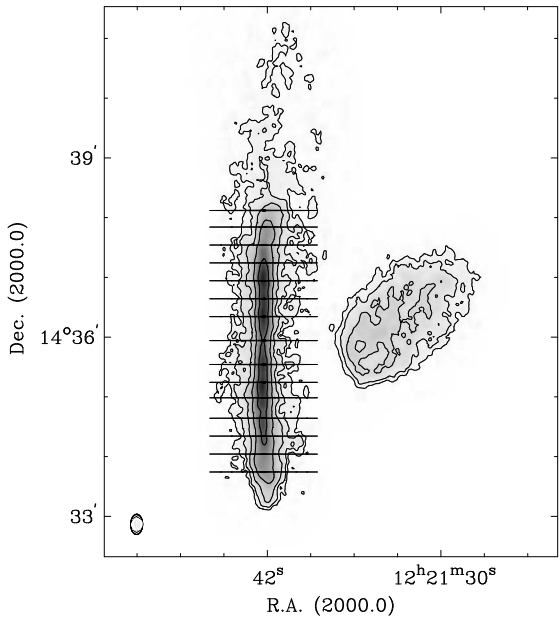}
\caption{\scriptsize\textit{A zeroth-moment map of NGC 4302 and its companion, NGC 4298. Contour levels begin at 1.8$\times$10$^{20}$ cm$^{-2}$ and increase by factors of 2.  The northern half is the approaching half.  Black lines represent slice locations and minor axis widths of bv diagrams in Figure~\ref{4302bvapproaching}.  Note the {\sc H\,i} tail, likely due to ram pressure stripping in NGC 4302.  Also note the emission in NGC 4298 that appears to be pushed in a similar direction, which would also be characteristic of ram pressure stripping.  The beam is shown in the lower left-hand corner. } \label{totalHIdata4302}}
\end{figure}

   Figure~\ref{totalHIdata4302} shows a zeroth-moment map of NGC 4302 and its companion, NGC 4298.  Note the tail on the northern half, which has been attributed to ram pressure stripping \citep{2007ApJ...659L.115C}. NGC 4298 shows asymmetries in a similar direction as the ram pressure stripped gas in NGC 4302 (i.e.\ the contours are closer together on the southeast edge than on the northwest edge), which may indicate ram pressure stripping in NGC 4298 as well.

\par
\begin{figure*}
\centering
\includegraphics[width=160mm]{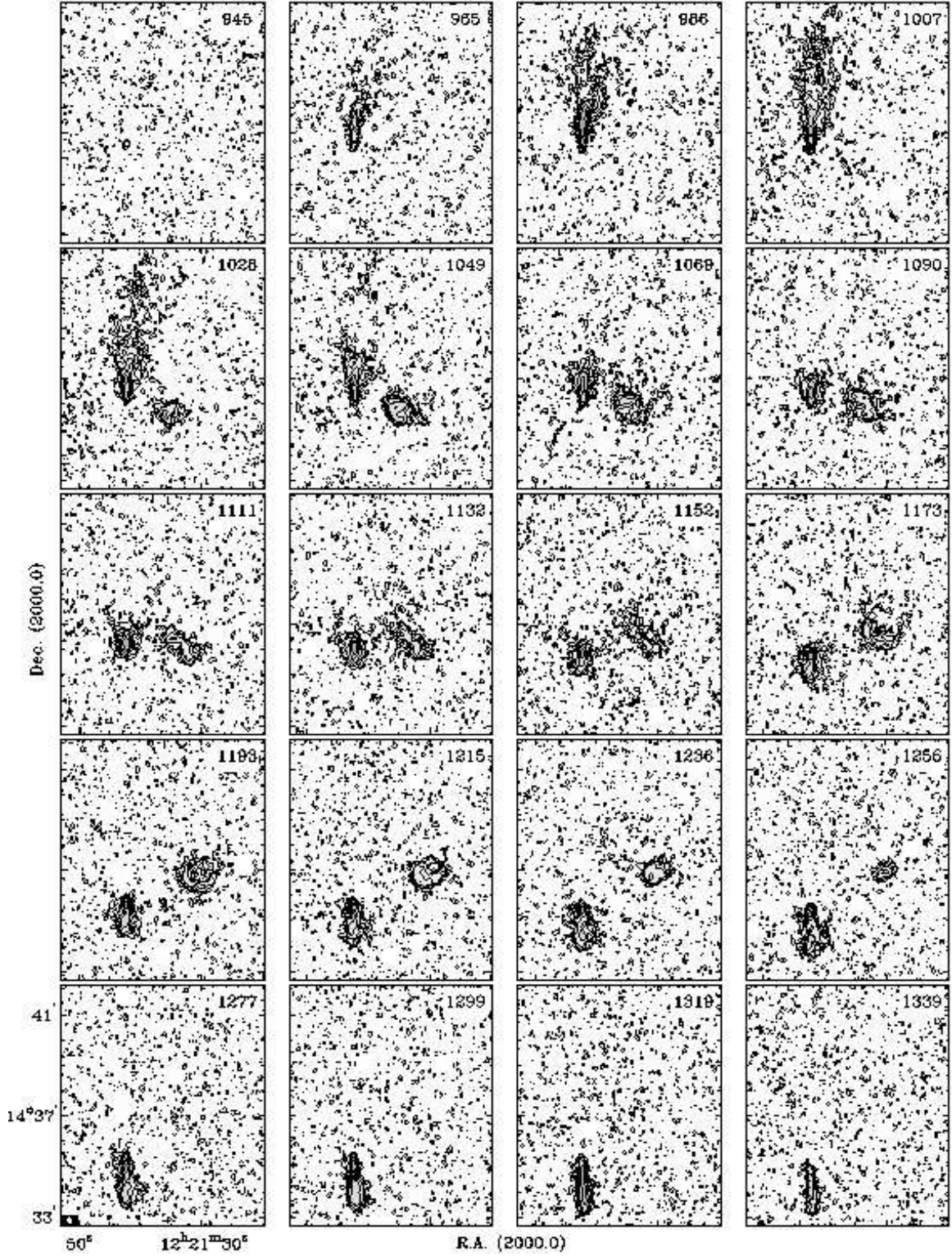}
\caption[Channel Maps of NGC 4302 and NGC 4298]{\scriptsize\textit{Channel maps of NGC 4302 and NGC 4298.  Velocities in km s$^{-1}$ are given in each panel.  Note the ram pressure stripping in the approaching half of NGC 4302, as well as the closely matching systemic velocities of the two galaxies. Contours begin at 2$\sigma$ (0.3 mJy bm$^{-1}$, or 7.2$\times$10$^{19}$ cm$^{-2}$) and increase by factors of two.} \label{channelmapall4302}}
\end{figure*}

   Figure~\ref{channelmapall4302} displays all of the channels from the cube that contain emission.  Note how closely the systemic velocities of both galaxies match (channels corresponding to 1132-1152 km s$^{-1}$).  Also note again the tail on the N (approaching) half. 

\begin{figure}
\centering
\includegraphics[angle = 270, width = 90mm]{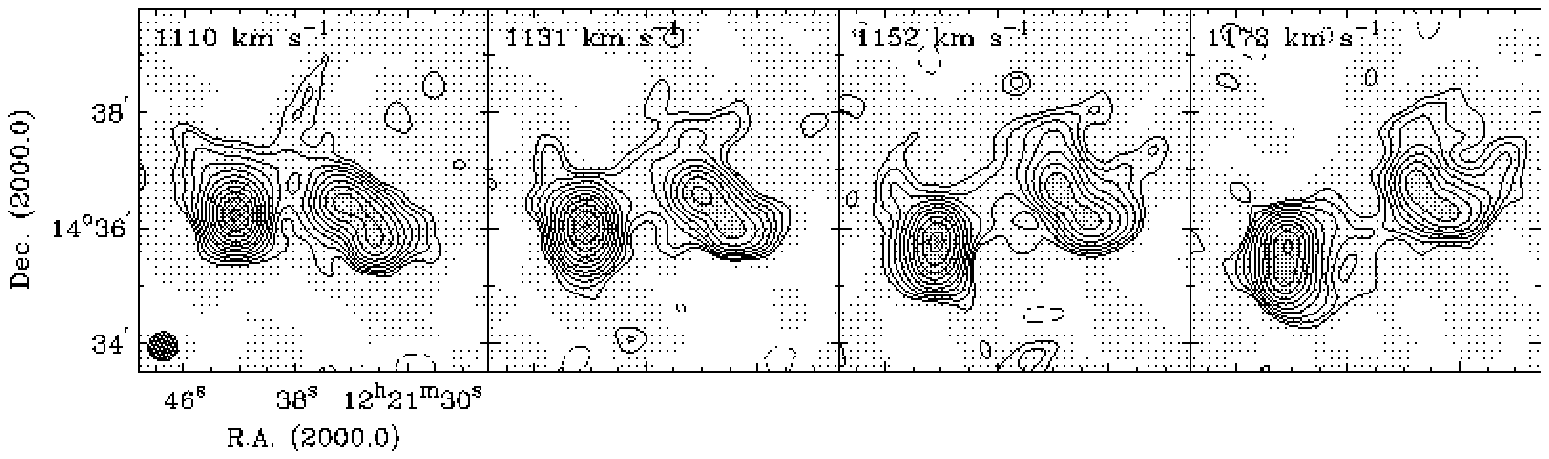}
\caption{\scriptsize\textit{Channel maps of a smoothed cube (30$\times$30") showing the most prominent section of the bridge between NGC 4302 and NGC 4298.  Velocities in km s$^{-1}$ are given in each panel.  Contours begin at 2$\sigma$ (0.048 mJy bm$^{-1}$) and increase by factors of $\sqrt{2}$.  The beam is in the lower left-hand corner of the first panel.} \label{channelmapbridge}}
\end{figure}

\par
   The galaxy is clearly asymmetric, with the ram pressure stripped gas evident on the approaching half, and various extra-planar clumps and extensions (e.g.\ the extension on the western side in channels with velocities of 1256-1299 km s$^{-1}$ in Figure~\ref{channelmapall4302}).  There is some unusual ``spiking" near the northern edges of channels between 1049 km s$^{-1}$ and 1132 km s$^{-1}$ (not seen in the figures presented here, but evident at low levels).  This spiking may be an artifact from the data reduction, but may also be related to the ram pressure stripping.   There are also indications of fluctuations in the position angle that are not characteristic of a warp (e.g.\ comparing the 1215 km s$^{-1}$ to 1319 km s$^{-1}$ channels, they show minor axis offsets in opposite directions from Dec. 14d 34m to 14d 36m rather than a general change in one direction), indicating an unusual degree of complexity in this galaxy.  Of course, such fluctuations are not necessarily due to changes in the orientation of the disk as a whole, but may be due to localized features, similar to the corrugations seen in {\sc H\,i} in other galaxies likely associated with global gravitational instabilities (e.g.\ \citealt{2008ApJ...688..237M}), or possibly associated with SF activity.  There are indications that NGC 4302 and NGC 4298 are interacting, best seen from channels 1110-1173 in Figure~\ref{channelmapbridge}.   This bridge is continuous in velocity, increasing the likelihood that it is a real feature.

\subsection{The Modeling of NGC 4302}

\par
   NGC 4302 is an abnormally complicated galaxy, which results in increased difficulty and uncertainties when creating tilted-ring models.  We attempt to minimize the addition of distinct features to our models.

\par
   We model the two halves separately.  The rotation curve is initially estimated using the values for the DIG presented in \citet{2007ApJ...663..933H}.  The surface brightness distribution is obtained using the {\tt GIPSY} task {\tt RADIAL}.  The central PA and inclination are assumed to be 180$^{\circ}$ and 90$^{\circ}$ respectively.  The velocity dispersion for most of the galaxy is modeled to be 30 km s$^{-1}$, which is unusually high, but may be related to the complex extra-planar {\sc H\,i} in this particular galaxy.  All of these quantities are adjusted throughout the modeling process. Systemic velocities between 1130 km s$^{-1}$ and 1180 km s$^{-1}$ are tested at early stages, with a value of 1150 km s$^{-1}$ used for most of the modeling.  Although, based on channel maps shown in Figure~\ref{channelmapall4302}, one may argue that a lower systemic velocity would be a better fit in central regions.  However, decreasing the systemic velocity would increase the discrepancy in the rotation curves in each half.  Thus, it may be that the inner and outer regions of NGC 4302 may have differing systemic velocities.  As mentioned earlier, such is also the case for NGC 3044, but in that galaxy, the overall shapes of the rotation curves on each side differed greatly, indicating that the velocities do in fact differ between the two halves,whereas in NGC 4302, the rotation curves are similar in shape if one excludes the ram pressure stripped gas.  Thus, the approach we use in NGC 4302 is to match the rotation curves to each other, which then dictates the central velocity.  The central position used in the models is R.A. 12h 21m 42.3s, Dec. 14d 35m 50.1s.

\subsubsection{Individual Models}
\par
\begin{figure}
\centering
\includegraphics[width=80mm]{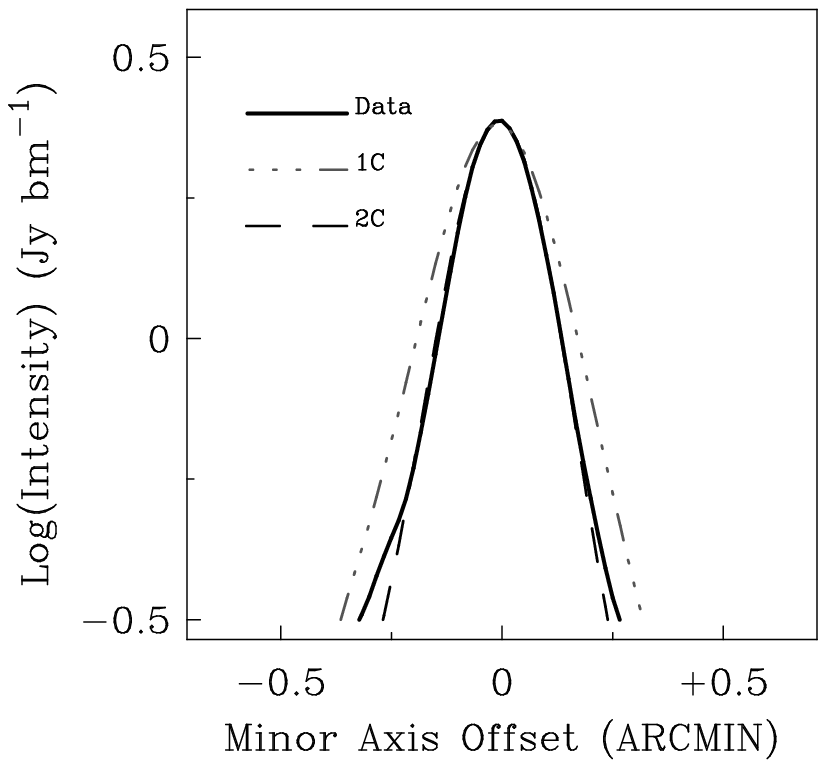}
\caption[Vertical Profiles of NGC 4302 and Models]{\scriptsize\textit{Vertical profiles of NGC 4302 (data), one-component (1C), and two-component (2C) summed over $\pm$40" (2.9 kpc) from the center. An alternative to the 1C model would be to decrease the scale height to fit the wings, which would consequently overshoot the peak in addition to not being as good a fit to other figures, thus such a model is omitted here. Note the substantial improvement gained by decreasing the height of the thin disk, and then adding a second, thicker layer with an hole near the center.} \label{vprofsumlog4302}}
\end{figure}
\begin{figure}
\centering
\includegraphics[width=80mm]{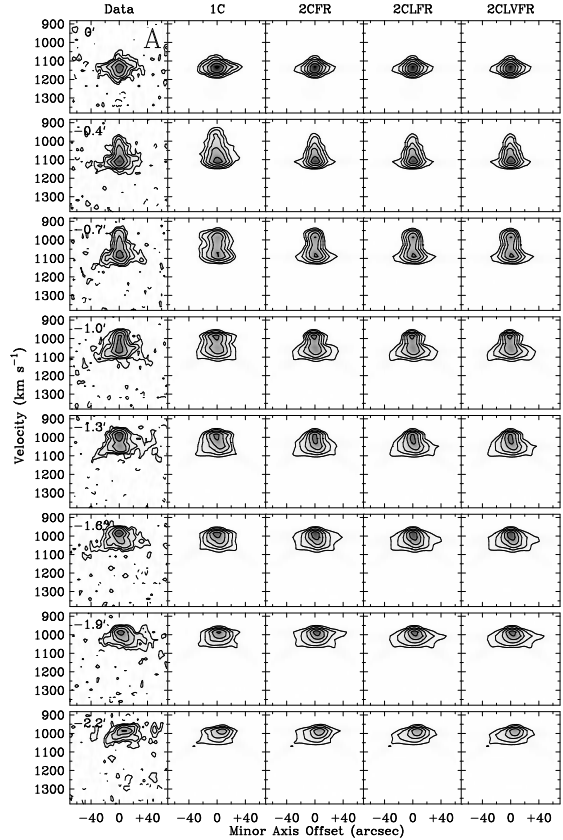}
\caption[Minor Axis Position-velocity Diagrams of NGC 4302 and Models]{\scriptsize\textit{Bv diagrams showing the approaching (A) and receding (b) halves of the data, 1C, 2CFR, 2CLFR, and 2CLVFR models of NGC 4302.  Slice locations are shown in Figure~\ref{totalHIdata4302} and positive minor axis offsets correspond to the western side. Notice the overall poor fit for the 1C model, especially the superfluous thickness near the center, largely on the terminal side.  The 2C model shows substantial improvement, in part due to the hole near the center, but also due to the increased spread in z, especially at large major axis offsets, especially on the systemic sides.  The 2CLFR and 2CLVFR models are the same for the approaching half, but differ slightly for the receding half.  The lag models are discussed further in $\S$~\ref{addlag}.  Contours are presented as in Figure~\ref{channelmapall4302}} \label{4302bvapproaching}}
\end{figure}

    Consistent with the approach to modeling NGC 3044, we again generate models of increasing complexity, starting with a single component (1C) model.  For such a model, we find an optimal scale height of 7.5" (540 pc).  As may be seen in the vertical profile shown in Figure~\ref{vprofsumlog4302}, the bv diagrams in Figure~\ref{4302bvapproaching}, and the channel maps in Figure~\ref{channelmapapproachingall4302}, not all the data can be fit by a single component alone.  Specifically, the 1C model is too thick in minor axis extent to match the data in the vertical profile, but if the scale height is decreased, the peak will rise too high above that of the data.  Additionally, the 1C model is too thick near the terminal sides of bv diagrams, especially near the center.  Again, in channel maps, the 1C model is too thick near the center.  Thus, we reduce the scale height of the first component, and add a second, thicker component with a hole near the center to the models.

\par
    A range of scale heights are explored, as well as different fractions of mass included in each component.  The final model consists of a thin disk having a scale height of 4" (300 pc), and a thick disk with a scale height of 25" (1.8 kpc).  Approximately 42$\%$ of the total mass resides in the thick component, but recall that the two components overlap at the midplane, thus not all of the emission in the thick component should be considered extra-planar.  Still, if one notes the substantial difference in scale heights of the two components, it is clear that most of the thick component resides well above the midplane.  (For reference, integrating the flux in the data itself shows that 28$\%$ of the total flux in all components is above a height of 1 kpc.)  The thick disk does not begin until a radius of 1.75' or 7.7 kpc (beyond this radius the two components share the same surface brightness distributions), effectively creating a high-z hole near the center.  This hole is necessary to fit the vertical profile, bv diagrams (Figure~\ref{4302bvapproaching}), and channel maps (Figure~\ref{channelmapapproachingall4302}) close to the center.  While we do not show models having high-z {\sc H\,i} near the center, one can imagine its inclusion would cause a thickening of the corresponding terminal sides of bv diagrams and central regions of channel maps not seen in the data.  Such a scenario may be remedied to some degree by adding a steep lag to the models, but attempts to do so result in extremely poor fits to the systemic sides of bv diagrams, and to the outer edges of channel maps (not to mention the poor fit to the vertical profile that eliminates such a model at an even earlier stage).

\begin{figure}
\centering
\includegraphics[width=80mm]{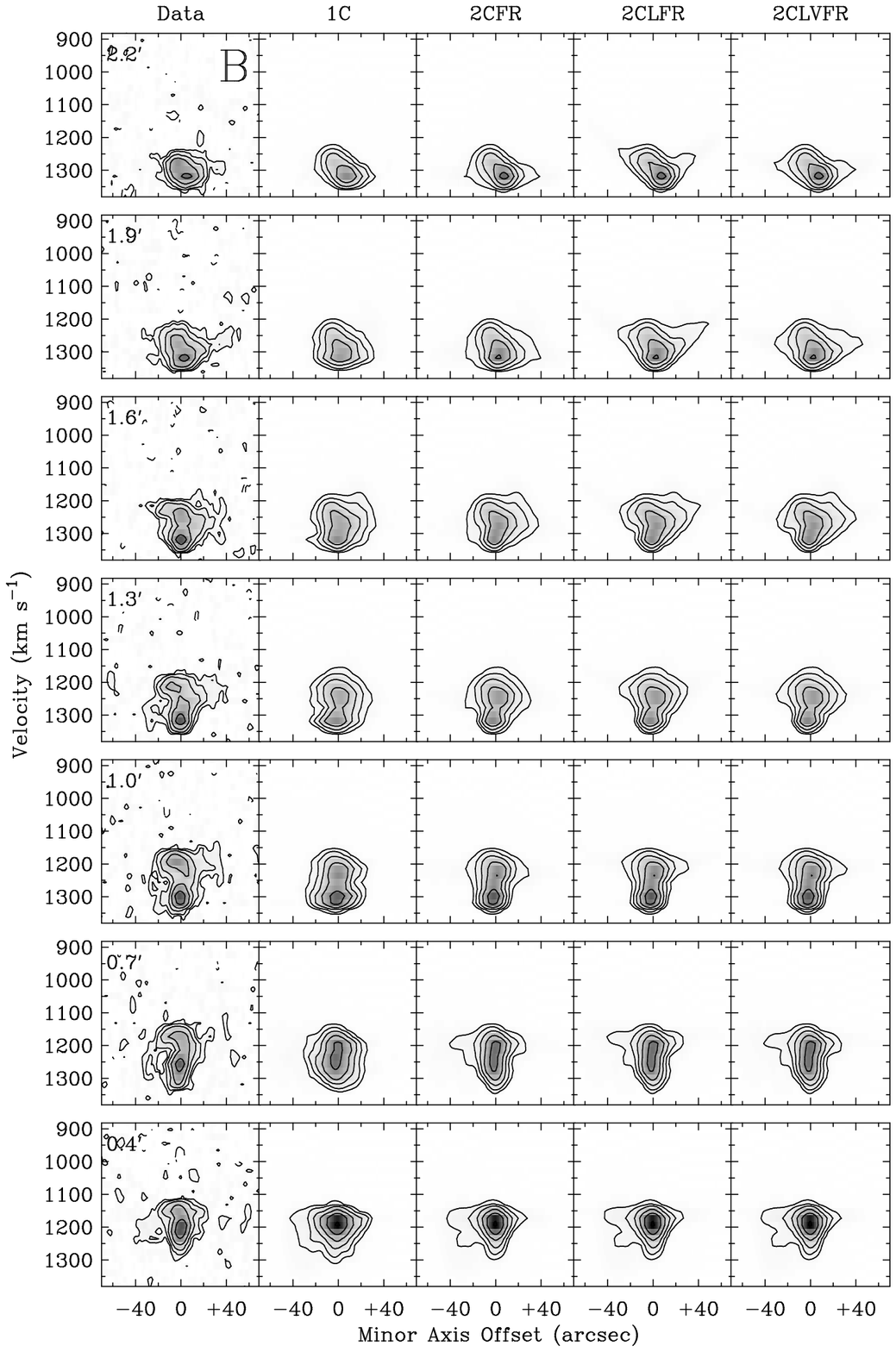}
\end{figure}

\begin{figure*}
\centering
\includegraphics[angle = 270, width = 160mm]{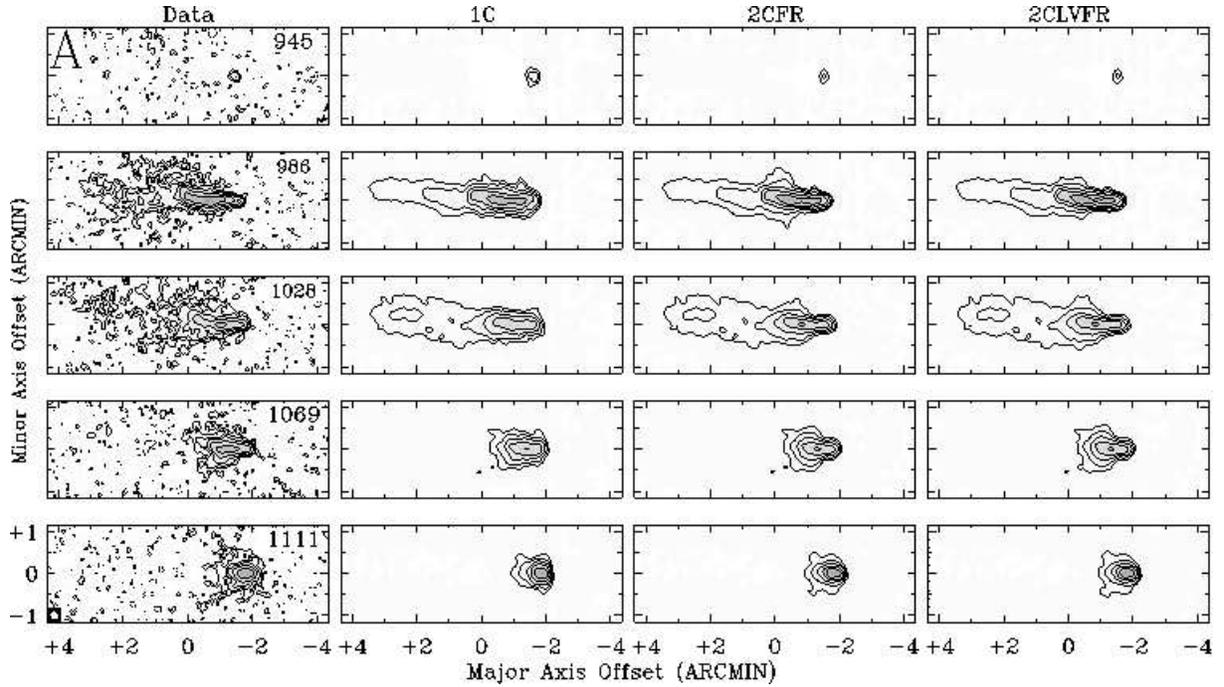}
\caption[Channel Maps of NGC 4302 and Models]{\scriptsize\textit{Channel maps showing the approaching (A) and receding halves (B) of NGC 4302.  Velocities in km s$^{-1}$ are given in each panel.  We model the ram pressure stripping in the approaching half (described in $\S$~\ref{distinct4302}).  Note the excess thickness near the center of the 1C model in both halves that is remedied by the 2CFR model (with a thinner main disk, and second, thicker component with a central hole).  The improvements due to the lag are minimal, especially in the approaching half, and are best seen in bv diagrams.  Velocities are given in each panel. Contours are presented as in Figure~\ref{channelmapall4302}.} \label{channelmapapproachingall4302}}
\end{figure*}

\begin{figure*}
\centering
\includegraphics[angle = 270, width = 160mm]{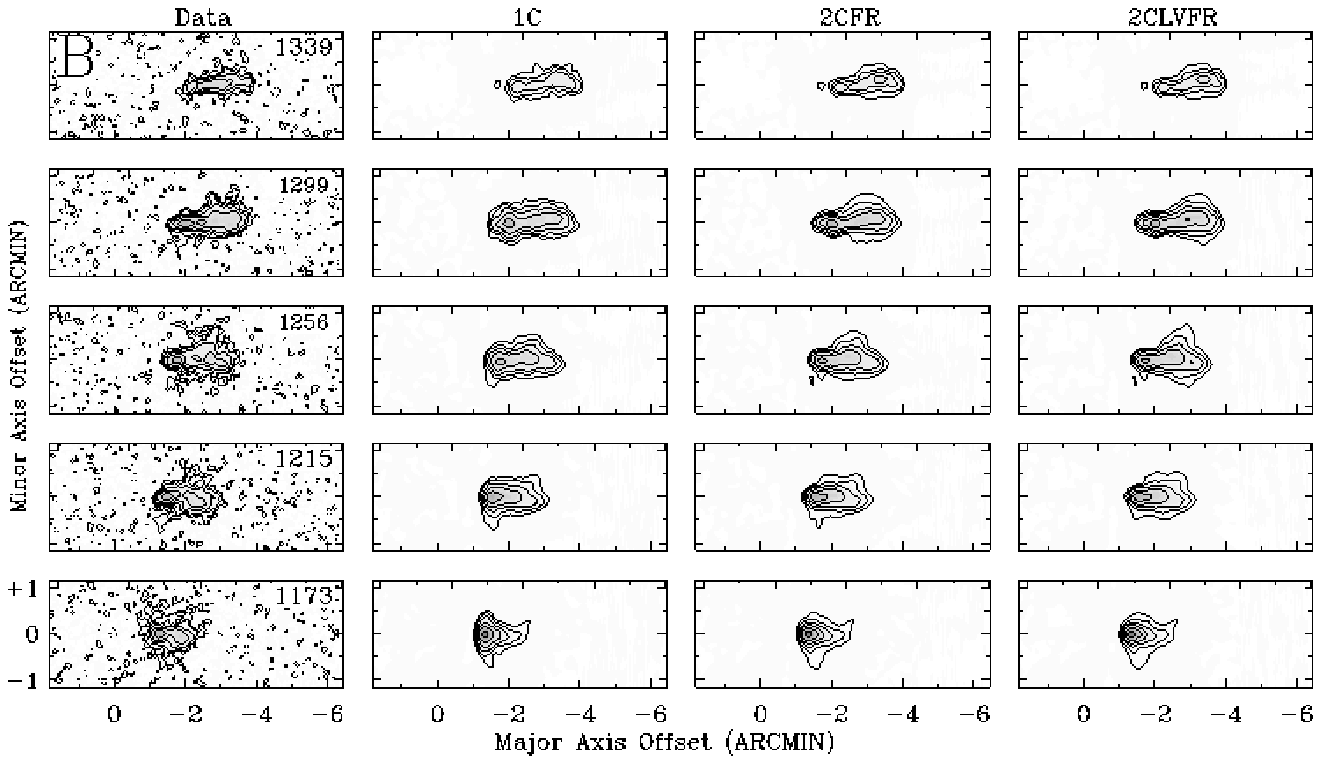}
\end{figure*}

\par
   Additionally, one could argue for a single, flaring layer instead of two components.  However, attempts to model a flare resulted in abrupt increases in the scale height rather than the gradual flares seen in other galaxies such as NGC 4565 \citep{2012ApJ...760...37Z}, indicating a fundamental difference between what is seen in NGC 4302 and a flare.  

\par
   We also consider a warp component along the line of sight as a means of thickening the disk but find that such a scenario would not be consistent with the data.  Specifically, we do not see the characteristic indentations on the systemic sides of bv diagrams (Figure~\ref{4302bvapproaching}), and the outer regions of channel maps (Figure~\ref{channelmapapproachingall4302}) that generally indicate a decrease in inclination.

\par
\begin{figure}
\centering
\includegraphics[width=80mm]{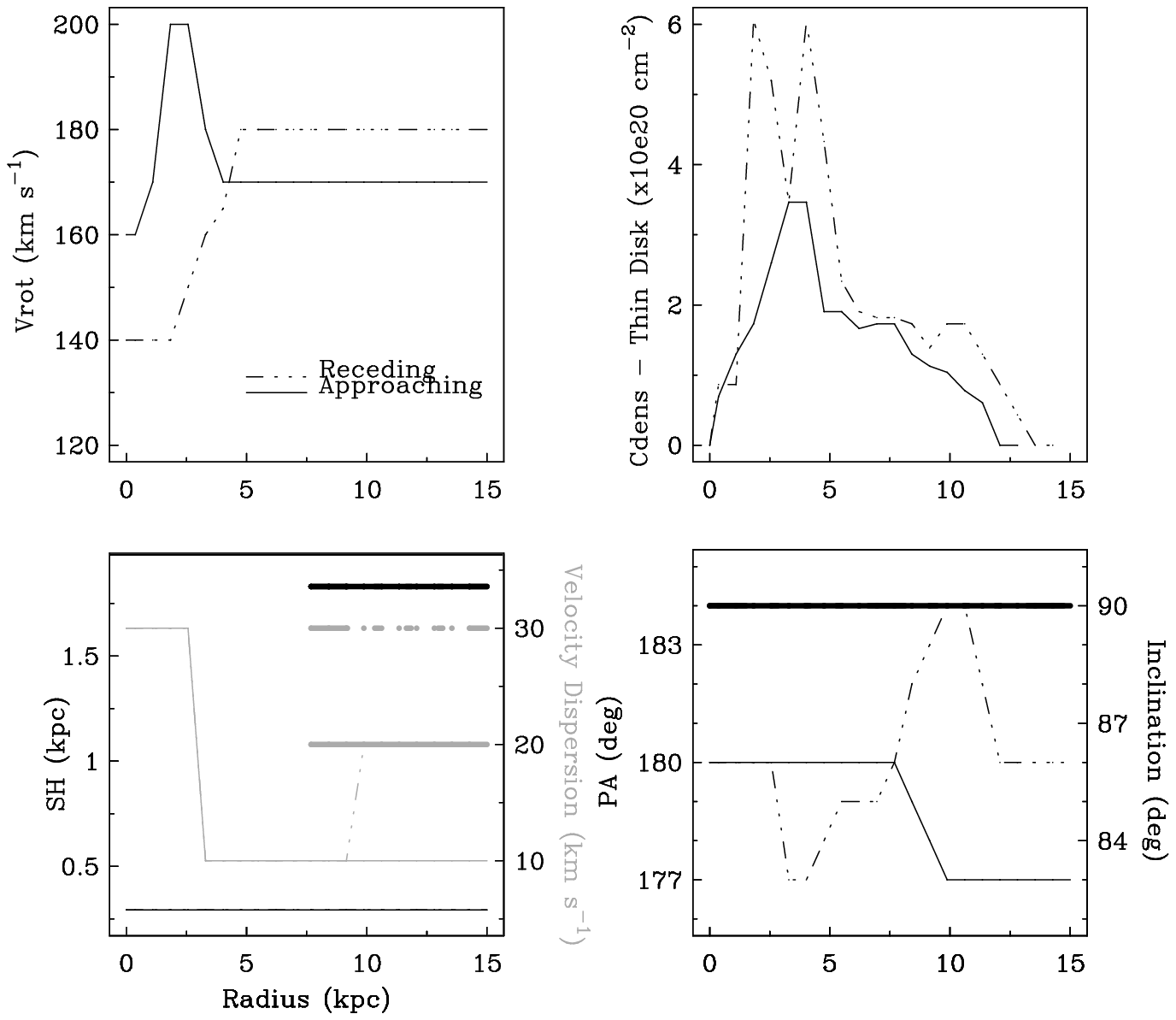}
\caption[NGC 4302 Model Parameters]{\scriptsize\textit{Parameters of the models for NGC 4302.  Note the difference in the rotation curves near the center.  Also note the fluctuations of the PA on the receding half. These fluctuations are not associated with any warping in NGC 4302, but are likely related to the distinct features described in $\S$~\ref{distinct4302}.  The inclination and scale height of the thick disk are plotted in bold while the velocity dispersion of all components are plotted in gray. The surface brightness is plotted for the thin disk, but is the same as the thick disk (aside from the hole, which ends at a radius of approximately 7.7 kpc).} \label{4302multiplot}}
\end{figure}

   Thus, we settle on a two component (2C) model with a high-z hole near the center.  Note even this model does not reproduce the wings at large offsets in Figure~\ref{vprofsumlog4302} due to distinct features and faint, asymmetric emission seen near the edges of the galaxy (Figure~\ref{totalHIdata4302}).  The parameters for the final model are shown in Figure~\ref{4302multiplot}.

\par
    Additionally, we see indications of variations in the velocity dispersion with radius shown in Figure~\ref{4302multiplot}.  We considered the most general trends for the velocity dispersion, prior to adding distinct features to the models, which will be discussed in $\S$~\ref{distinct4302}.  While we avoid allowing more localized features to impact properties of the main disk as much as possible, it may be that some of the changes in the velocity dispersion are related to these features.

\subsubsection{Distinct Features Included in the Models}\label{distinct4302}
\par
\begin{figure}
\centering
\includegraphics[width=80mm]{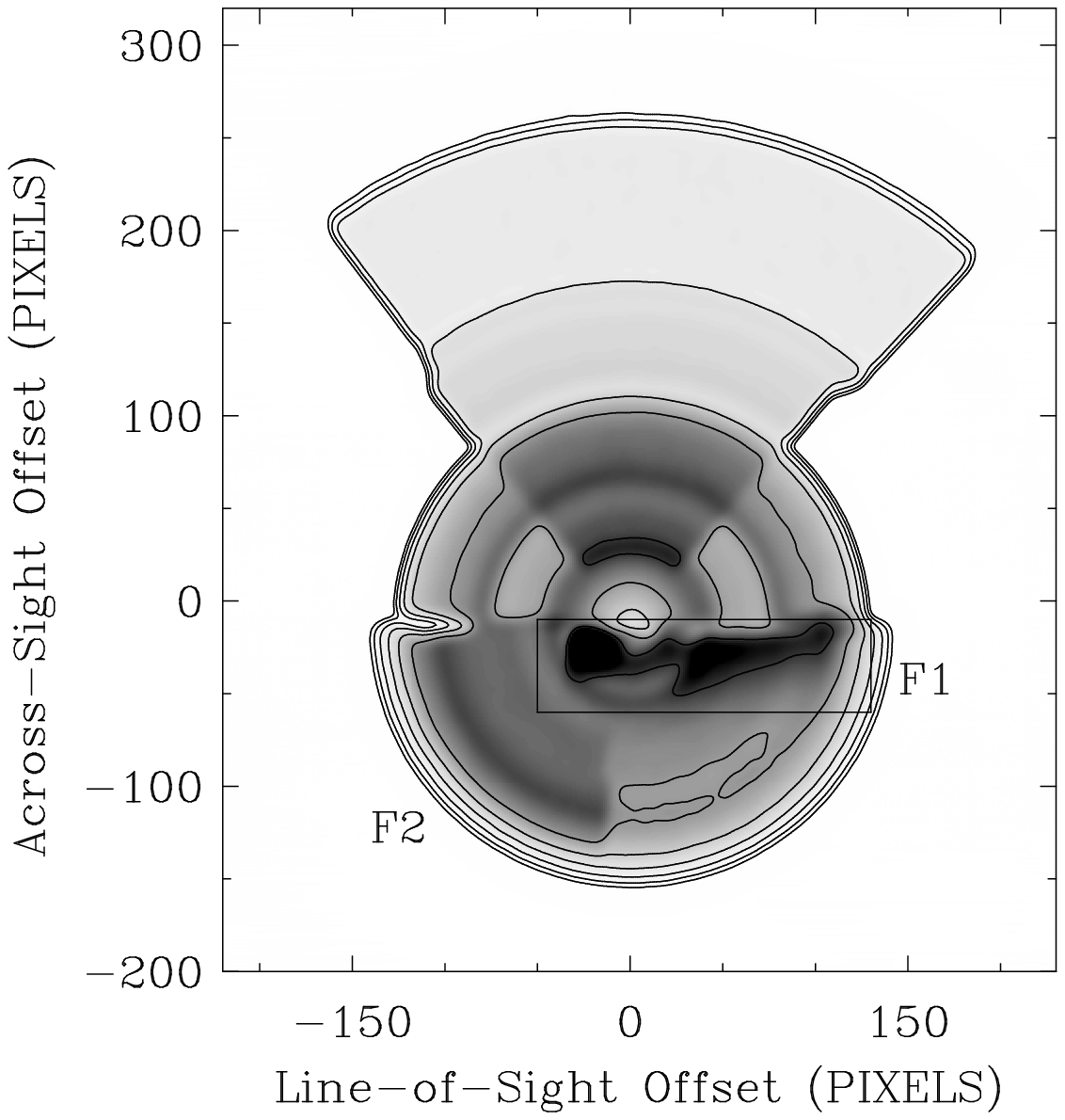}
\caption[]{\scriptsize\textit{A face-on representation of the optimal model including the ram pressure stripped gas (top wedge) and Features 1 and 2 (described below) in NGC 4302.  Feature 1 is enclosed by the box labeled ``F1" and Feature 2 is the wedge centered azimuthally on the ``F2" label.  Pixels are equivalent to 1.4" (100 pc). The observer is located on either the left or right of the center of the galaxy (due to ambiguities introduced by projection effects).} \label{totalHIface4302}}
\end{figure}

   While there is much anomalous gas in NGC 4302, we attempt to model only three distinct features (2CFR models include these features and their associated radial motions).  Additionally, although not really ``features" per se, we add fluctuations to the PA on the receding half, and fluctuations in the velocity dispersion on both halves.  \textit{Unlike the case of NGC 3044, these features affect lag values.} With their inclusion, we can say something, however minor, about the kinematics of this gas while making no claims about its origins.  Although they are added to the models at the same stage, we cannot claim how closely or even whether they are associated with each other.  Except for the ram pressure stripped gas, we do not assign physical meaning to these features - our primary goal is to assess how much they impact lag determinations.  They are included in all models unless otherwise specified.  A face-on representation of these features is given in Figure~\ref{totalHIface4302}.

\par
\begin{figure}
\centering
\includegraphics[width=80mm]{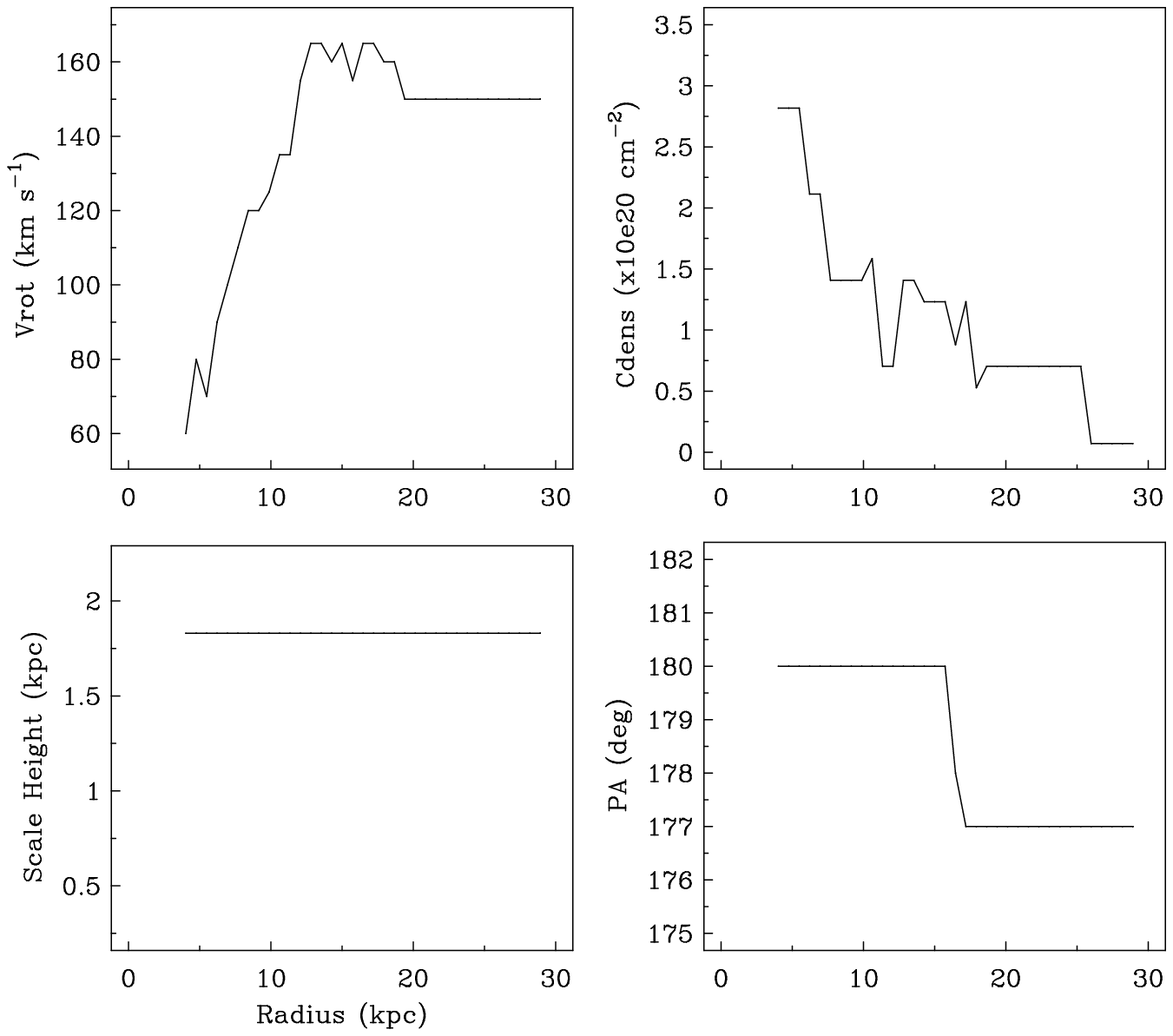}
\caption{\scriptsize\textit{The model parameters for the ram pressure stripped gas in NGC 4302.  Note the large radius to which the stripped gas extends, as well as the overlap with the main disk.  The apparent rotational velocities of the ram pressure stripped gas in the region in which the stripped gas overlaps with the main disk are slower than those of the main disk in our models.  However, due to the combined effects of the two model components, and likely peculiar motions in the stripped gas, these velocities should not be interpreted as true rotational velocities.} \label{4302parameterram}}
\end{figure}

   Immediately clear from channel maps (Figure~\ref{channelmapall4302}) is the ram pressure stripped gas on the approaching (N) half.  While ram pressure stripping by nature is not modeled well by concentric rings, we attempt to approximate its morphology and kinematics through a wedge (i.e.\ a feature with limited radial and azimuthal extent in the disk).  This wedge begins at a radius of approximately 1' (4.0 kpc) on the approaching half, and extends to 6.4' (28.2 kpc).  The azimuthal width of each arc is 80$^{\circ}$, and each is azimuthally centered on the approaching half.  The velocity dispersion is 30 km s$^{-1}$, and the inclination is 90$^{\circ}$. The ram pressure stripped gas is best modeled with a lag of 21 km s$^{-1}$ kpc$^{-1}$ beginning 18" (1.3 kpc) above the midplane.  The corresponding surface brightness distribution, PA distribution, rotation curve, and scale height are given in Figure~\ref{4302parameterram}.  Discussion of the validity of the assumptions used in our models regarding this gas is provided in $\S$~\ref{additional4302}.

\par
   There exists a disturbed region in the receding half of NGC 4302 from a radius of 0.35' (1.5 kpc) to 2.6' (11.2 kpc).  In order to better fit this region, we add two more features to our models.

\par
\begin{figure*}
\centering
\includegraphics[angle = 270, width = 150mm]{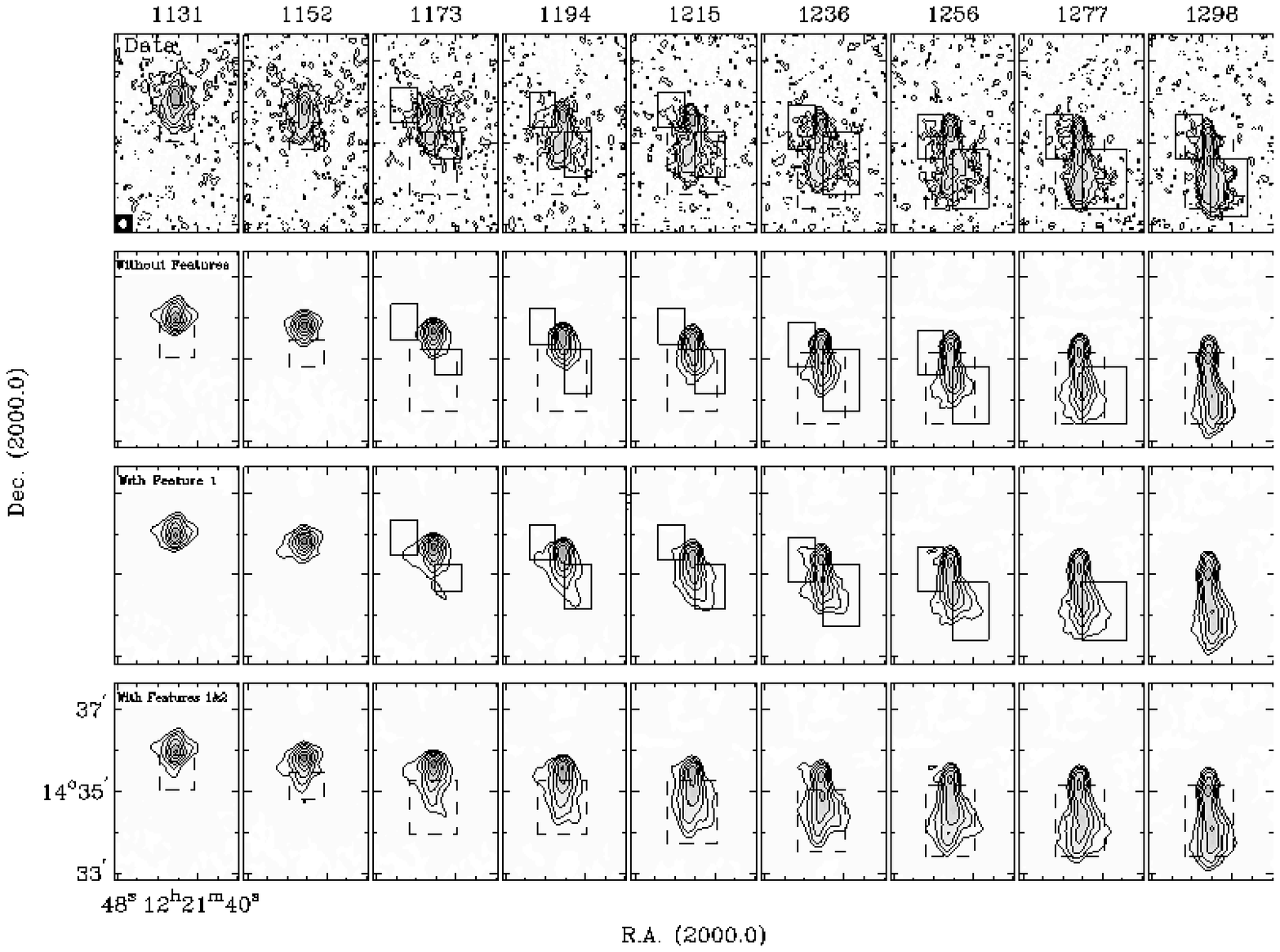}
\caption{\scriptsize\textit{Channel maps showing the data, and models with and without ``Features 1 and 2" of NGC 4302.  Velocities in km s$^{-1}$ are given above each column. Boxes with solid lines highlight the regions in which improvement is seen through the addition of Feature 1, while dashed boxes indicate regions of improvement due to Feature 2.  It should be noted that the fit is imperfect, but the features improve the match between the model and data. Velocities are given above each column. Contours are presented as in Figure~\ref{channelmapall4302}.} \label{channelmapfeatures}}
\end{figure*}

   One may first note a protrusion in the SW quadrant of the galaxy (Feature 1, Figure~\ref{channelmapfeatures}, lower-right boxed region).  This feature extends from velocities 1173 km s$^{-1}$ to 1298 km s$^{-1}$.  A seemingly un-related protrusion can also be seen in the NE quadrant of the receding half in this velocity range (upper-left boxed region).  At first we attempt to model these features separately, but then note the slant of extra-planar emission in bv diagrams from 0.4' to 1.9'.  This slant is best seen at 1.3' in Figure~\ref{4302bvapproaching} where an excess of extra-planar emission can be seen on both the positive and negative minor axis off-sets, with the emission on the negative side being closer to the terminal velocities, and the emission on the positive side closer to the systemic.  The 1.6' and 1.9' panels show extra-planar emission primarily on the positive side.  Judging by this behavior, it appears as though the two protrusions could in fact be related.  Thus, we model them as as a single, elongated, curved feature, running at an angle to both the planes of the galaxy and the sky, with a gaussian intensity profile along its short axis with a systemic velocity of 1250 km s$^{-1}$, a rotational velocity of 20 km s$^{-1}$, and centered at RA 12h 21m 44.9s and Dec. 14d 35m 42s.  This systemic velocity was chosen based on the velocity at which the emission appears to be centered between the two protrusions.  We do not mean to suggest that this feature is rotating around a different kinematic center than is the disk (although such could be the case), but chose these parameters as a convenient way to model the feature.  The amplitude of the gaussians varies along the curve, as does the velocity dispersion (starting at 80 km s$^{-1}$ for the first half (radially), and then linearly decreasing to 60 km s$^{-1}$.  It should be noted that it is possible that these high values for the dispersion parameter may represent a spread of line-of-sight velocities rather than true dispersions. The tilt of the feature causes slanting from the box in the upper left-hand corner, down to the box in the lower right-hand corner in Figure~\ref{channelmapfeatures}.  The curved feature intersects the disk, and has a substantial line of sight extent.

\par
    A second feature is added to the models after it is seen that channels in Figure~\ref{channelmapfeatures} between 1152 km s$^{-1}$ and 1298 km s$^{-1}$ in the model are missing flux relative to the data at intermediate and large radii (indicated by dashed boxes in Figure~\ref{channelmapfeatures}).  Increasing the surface brightness for entire rings, or changing the velocities, does not remedy this issue without compromising the fit for other channels or different radii.  Thus, we add a wedge centered at an azimuthal angle of 135$^{\circ}$, comprised of arcs having widths of 80$^{\circ}$, to increase the flux in this region only.  The total velocity dispersion of this feature is 20 km s$^{-1}$.  The feature extends from 6 to 12 kpc, with a column density of approximately 2$\times$10$^{20}$ cm$^{-2}$.  The rotational velocity rises from 85 km s$^{-1}$ at a radius of 6 kpc to 150 km s$^{-1}$ at 10 kpc.  This feature is denoted as ``Feature 2" in figures and throughout the text.  Potential origins of both Feature 1 and Feature 2 will be discussed in $\S$~\ref{additional4302}.  The improvement in the channel maps due to these features is clear, by examining the 3rd and 4th rows of Figure~\ref{channelmapfeatures}.

\par
  Fluctuations in PA were added at small and moderate radii on the receding half in combination with the features mentioned above to smooth the transition between the main disk and the anomalous gas.  It should be noted that, although adjustments to the position angle are typically associated with a warp component across the line of sight, these local fluctuations \textit{should not} be considered directly related to any warping of the galaxy.

\subsubsection{The Addition of a Lag}\label{addlag}

\begin{figure}
\centering
\includegraphics[width=80mm]{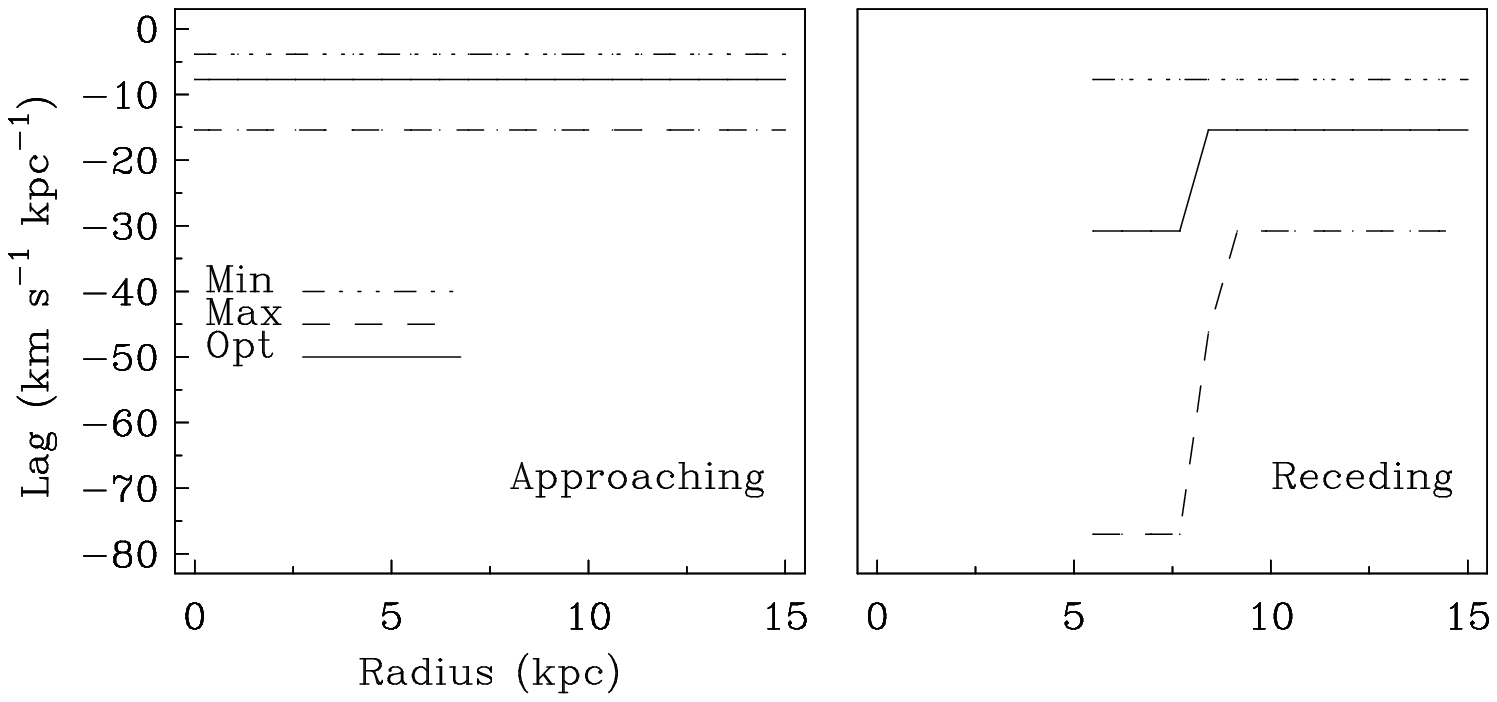}
\caption{\scriptsize\textit{The lag values and their uncertainties for the models with lags beginning at the midplane in NGC 4302.  A radial shallowing is detected on the receding half.} \label{parameterslag4302midplane}}
\end{figure}

\begin{figure}
\centering
\includegraphics[width=80mm]{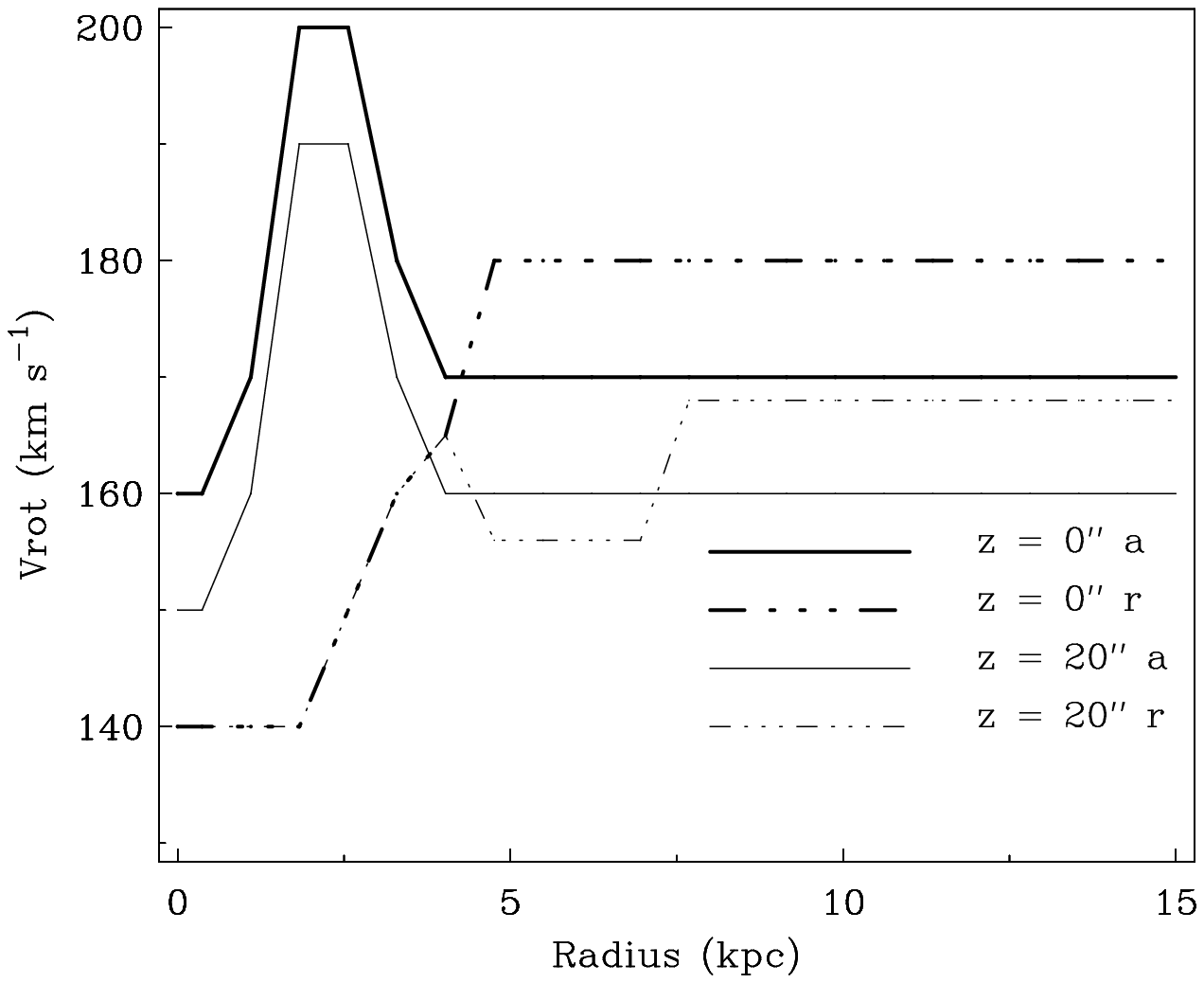}
\caption{\scriptsize\textit{The rotation curves at the midplane and at z=20" for the models having lags that start at the midplane in NGC 4302.  The approaching half is indicated by ``a" and the receding half is indicated by ``r."} \label{4302parametervrotsmidplane}}
\end{figure}

\par
   Judging by bv diagrams and channel maps (Figures ~\ref{4302bvapproaching} and~\ref{channelmapapproachingall4302}), the two component model is still not an ideal fit to the data.  Emission in the data is effectively displaced toward the systemic side with increasing minor axis off-set in the data in bv diagrams relative to the 2C model, where there is a surplus of emission near the terminal side (best seen beyond major axis offsets of $\pm$1.6').  To remedy these issues, we now add a lag (2CLFR) to the models.  Note that these models already include a second component (2C) and additional features and radial motions (FR).

\par
    The optimal model for a lag beginning at the midplane would have a lag of $-$7$^{+3}_{-7}$ km s$^{-1}$ kpc$^{-1}$ on the approaching half, and $-$27$^{+21}_{-55}$ km s$^{-1}$ kpc$^{-1}$ on the receding half. Figure~\ref{parameterslag4302midplane} shows the lag values and their uncertainties for this model, and Figure~\ref{4302parametervrotsmidplane} shows the resulting rotation curves at high z for shallower lags beginning at the midplane.

\par
   Because of the apparent flattening of the terminal sides in some bv diagrams (e.g.\ panels at -1.3' and -1.6' in Figure~\ref{4302bvapproaching}), we have explored the possibility of a lag beginning higher above the midplane, although we do not present such a model here due there not being enough marked improvement to warrant the increase in free parameters.  We make note of such a possibility here because this is the one galaxy in our sample which does not show conclusive evidence of a lag beginning close to the midplane.

\par
   It should be noted that the radii at which the lags begin are dictated by when the lag begins to produce some effect, not because a lag within those radii is detrimental.  This should be unsurprising given the extra-planar hole near the center that extends to just beyond the radius at which lag effects are noticeable.

\par
    Given the tumultuous nature of NGC 4302, a lag is difficult to constrain.  Qualitatively, it is reasonably clear by the emission in bv diagrams that is displaced toward the systemic at high-z (Figures~\ref{4302bvapproaching}) that gas at high-z is rotating at a slower rate than gas at the midplane.  Additionally, as in other galaxies, judging by the more rounded as opposed to v-shaped terminal sides of bv diagrams at large radii in the receding half (Figure~\ref{4302bvapproaching}, 1.9' and 2.2'), the lag appears to shallow with radius in the receding half (2CLVFR).  However, the extra-planar kinematics in general are likely impacted by ram pressure stripping and interactions with the companion galaxy.  The additional features described in $\S$~\ref{distinct4302} were found to impact values in our models, resulting in a compromised ability to determine reliable values for NGC 4302 as compared to other galaxies.

\subsubsection{The Final Models and Their Uncertainties}\label{uncertainty4302}
\par
\begin{figure}
\centering
\includegraphics[width=80mm]{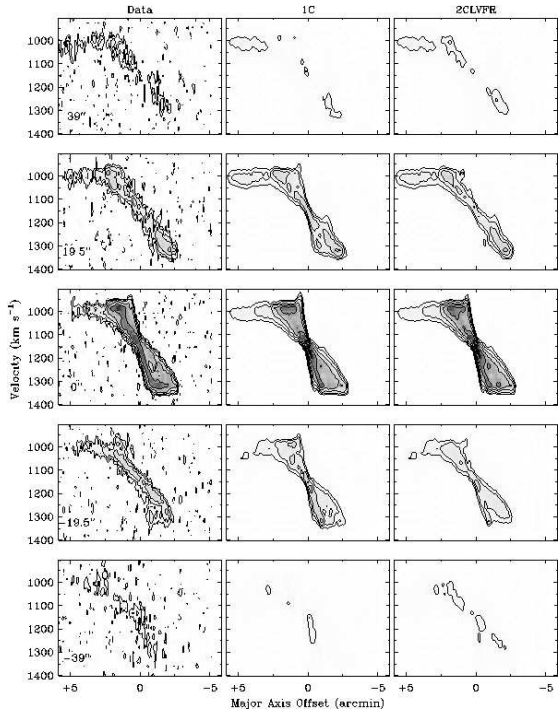}
\caption{\scriptsize\textit{Lv diagrams showing the data, 1C, and 2CLVFR models for NGC 4302.   Minor axis offsets are given in each panel and positive offsets correspond to the western side. Most improvements are due to the addition of a second, thicker component.  Note the improved fit to the panels high above the midplane.  Contours are presented as in Figure~\ref{channelmapall4302}.} \label{4302lv}}
\end{figure}

   We will now consider a comparison of the sum of the square of the residuals for each model.  There is a 17.8$\%$ decrease between the 1C and 2CFR models.  There are slightly larger decreases of 18.1$\%$ and 18.3$\%$ for the 2CLFR and 2CLVFR (final) models.  The overall trend is for sum of the square of the residuals to decrease with each improved model, but by far the largest change is from adding a second component.  All of these models include the extra features described in $\S$~\ref{distinct4302}.  Lv diagrams of final models data, 1C and final models are shown in Figure~\ref{4302lv}.

\par
   Now we consider the sum of the square of the residuals in the context of the additional features added to the models.  The percent decrease when adding the ram pressure stripped gas to the 2CLV model (a two component model with a radially shallowing lag but no additional features or radial motions, not shown in figures) with a lag starting high above the midplane (other features are still included) is 15.6$\%$.  Adding Feature 1 results in a decrease of 3.3$\%$, and adding Feature 2 results in a decrease of 10.0$\%$.  These numbers indicate that the ram pressure stripped gas and Feature 2 improve the models substantially, while effects from Feature 1 are minimal.  However, it should be noted that the improvement from any one of these features yields a greater decrease than the addition and/or refinement of the lag.  

\subsection{A Brief Summary of the Properties of NGC 4298}
\par
    We model the companion galaxy of NGC 4302, NGC 4298.  The position angle, inclination, and kinematic center are estimated by eye, and initially refined by tilted ring fitting using the {\tt GIPSY} task {\tt ROTCUR}.   Also using {\tt ROTCUR}, we estimate the systemic velocity (1140 km s$^{-1}$) and rotation curve.  We use {\tt ELLINT} to estimate the surface brightness distribution.  These quantities are later refined in {\tt TiRiFiC}.  The final value of the position angle is 313$^{\circ}$.  The inclination is 70$^{\circ}$, and the kinematic center is RA 12h 21m 32.6s, Dec. 14$^{\circ}$ 36' 22.7".  The rotation curve peaks at 110 km s$^{-1}$ and holds constant before decreasing at a radius of 1.4' (6.2 kpc).  The total {\sc H\,i} mass we estimate is 6.3$\times$10$^{8}$ $M_{\odot}$.  A rough estimate for the dynamical mass of NGC 4298 based on the kinematics we model is 1.5$\times$10$^{10}$ $M_{\odot}$, while NGC 4302 is 1.0$\times$10$^{11}$ $M_{\odot}$.  Of course, these dynamical masses are lower limits, even more so for a moderately inclined galaxy for which we cannot detect a given distribution of {\sc H\,i} to as large a radius.  Concerning extra-planar emission, we see no clear evidence for ``beard" gas in NGC 4298, although the ram pressure stripping could have disturbed any that would have been present.

\section{Discussion}\label{discussion}
\subsection{NGC 3044}
\subsubsection{The Lag in NGC 3044}
\par
    The lag in NGC 3044 is relatively steep compared to galaxies such as NGC 4244 and NGC 891 (Table~\ref{table_5}).  It starts close to the midplane and shallows radially.  In the approaching half, the shallowing begins near a radius of 1' (5.4 kpc), and near 1.4' (7.6 kpc) in the receding. Both are well within $R_{25}$ (2.5' or 13.6 kpc).  The lags reach zero close to $R_{25}$ itself.  The possible relevance of this result will be discussed in $\S$~\ref{radialvariations}.

\subsubsection{A Comparison to Previous Work}
\par
     As mentioned previously, \citet{1997ApJ...490..247L} observed and modeled the {\sc H\,i} in NGC 3044.  A primary focus of their paper was localized and distinct extra-planar features, which we consider only briefly.  They do not consider vertical gradients in rotation speed, which is the core of our work.  As for the models themselves, the inclinations found by both groups agree (85$^{\circ}$).  Adjusting for the difference in distances used, the exponential scale height they determine (490 pc) is somewhat narrower than ours (635 pc).  For the kinematic center, we find a difference in right ascension of 0.6s, and a difference in declination of 8.6", or approximately 12" (1.1 kpc) total between our models. Perhaps the greatest difference in the models is the systemic velocity.  They find a systemic velocity of 1287 km s$^{-1}$, while we find 1260 km s$^{-1}$.  Related to this, the rotation curve modeled by \citet{1997ApJ...490..247L} has approximately equal average rotation speeds in each half, while ours differs by approximately 30 km s$^{-1}$ between the two.  If we change the systemic velocity to 1280 km s$^{-1}$ in our models, the two halves have approximately equal average rotation speeds.  We choose our final systemic velocity by matching the central channels (Figure~\ref{3044channelmapcenter2}), and then changing the rotation curve to accommodate.  Although we are confident in the value determined using our models, the absolute value of the systemic velocity has no impact on lag values.

\subsection{NGC 4302}
\subsubsection{The Interpretation of Additional Features in the Models}\label{additional4302}

\par
   We model ram pressure stripped gas using azimuthal arcs in one half of the galaxy.  This method can only be used as a very loose approximation.  The arcs in our models create a wedge centered on the approaching half, although both halves are likely affected. 

\par
     It is unlikely that the ram pressure stripped gas would retain the same systemic velocity as the rest of the galaxy.  We do not incorporate this aspect into our models as a separate systemic velocity for the ram pressure stripped gas could not be reliably constrained.  We obtain a rough approximation of the rotation curve of the stripped gas that is significantly slower than the main disk where they overlap.   Outside the region of overlap, the rotation curve of the stripped gas is approximately 20 km s$^{-1}$ slower than the peak rotational velocity of the disk.

\par
   Features 1 and 2 could be related to merger activity, ram pressure stripping, spiral structure, disk-halo flows, or a combination of these phenomena.  The location and velocities of Feature 1 in relation to the companion galaxy indicate that it may be due to an interaction between the two galaxies.  The channels in which the companion is closest to NGC 4302 spatially are associated with the approaching half of the galaxy (channels having velocities of 1069-1152 km s$^{-1}$).  Thus, the slower peculiar velocity of this feature may indicate accretion of material from the companion or tidal interactions.  It is also possible that Feature 1 could be due to disk-halo flows.  In contrast, Feature 2 is less likely to be due to disk-halo flows or merger activity, but more likely to be due to spiral structure.  It is difficult to even speculate on the degree to which ram pressure stripping could affect either feature.

\subsubsection{The Lag in NGC 4302}
\par
   Due to the unusually complicated nature of NGC 4302, it is extremely difficult to constrain its lag.  Attempting to do so pushes the limits of tilted-ring modeling, and limited weight should be given to results for this galaxy.

\par
   It is possible that the lag in NGC 4302 may start above the midplane.  We do not see such behavior in the other galaxies included in our sample, which makes this characteristic somewhat intriguing.  

\par
   Among the foremost reasons for the difficulty in constraining the lag is the possible interaction between NGC 4302 and its companion.  There are indications that a bridge between the two galaxies is likely present (Figure~\ref{channelmapbridge}, especially in channels including and between 1110 km s$^{-1}$ and 1173 km s$^{-1}$).  Furthermore, their small spatial separation in their projected region of closest approach ($\leq$0.5' or 2.2 kpc) and shared velocity range (the systemic velocity of the companion is close to that of NGC 4302) increase the likelihood of the companion having some bearing on the kinematics of NGC 4302.  The region encompassing Feature 1 may indeed be strongly affected by an interaction, and possibly accretion from the companion galaxy.  The velocities of both Feature 1 and Feature 2 are peculiar, with those of Feature 2 clearly slower than that of the main disk of NGC 4302.

\par
   In spite of the issues mentioned above, the lag in the receding half shallows and reaches its lowest value just within the optical radius. 
\subsubsection{A Comparison to Previous Work}
\par
    \citet{2007ApJ...663..933H} found a DIG lag of $-$39 km s$^{-1}$ kpc$^{-1}$ in the NW quadrant of NGC 4302, and $-$25 km s$^{-1}$ kpc$^{-1}$, \textit{steepening} to $-$65 km s$^{-1}$ kpc$^{-1}$ near a radius of 3.8 kpc in the SW quadrant (using our assumed distance).  

\par
   The lag of $-$7$^{+3}_{-7}$ km s$^{-1}$ kpc$^{-1}$ that we detect in the approaching half is significantly shallower than the DIG lag of $-$39 km s$^{-1}$ kpc$^{-1}$.  Some of the difference may possibly be attributed to ram pressure stripping, which, if all else is equal, would have a greater effect on the DIG kinematics than those of the {\sc H\,i} simply because DIG typically has lower density than neutral gas.  However, much of the DIG is concentrated near the center of NGC 4302, which would render it more gravitationally bound than {\sc H\,i} at larger radii.

\section{Trends in Lags Among Galaxies}

\par
    As described in $\S$~\ref{introduction}, other extra-planar phases are likely related to SFR on both global and localized scales.  However, no such relation has been established for extra-planar {\sc H\,i}.  We defer an analysis of whether morphological properties of extra-planar {\sc H\,i} relate to SFR to a future paper from the HALOGAS project and focus here on kinematics.  

\par
\begin{figure*}
\centering
\includegraphics[width=160mm]{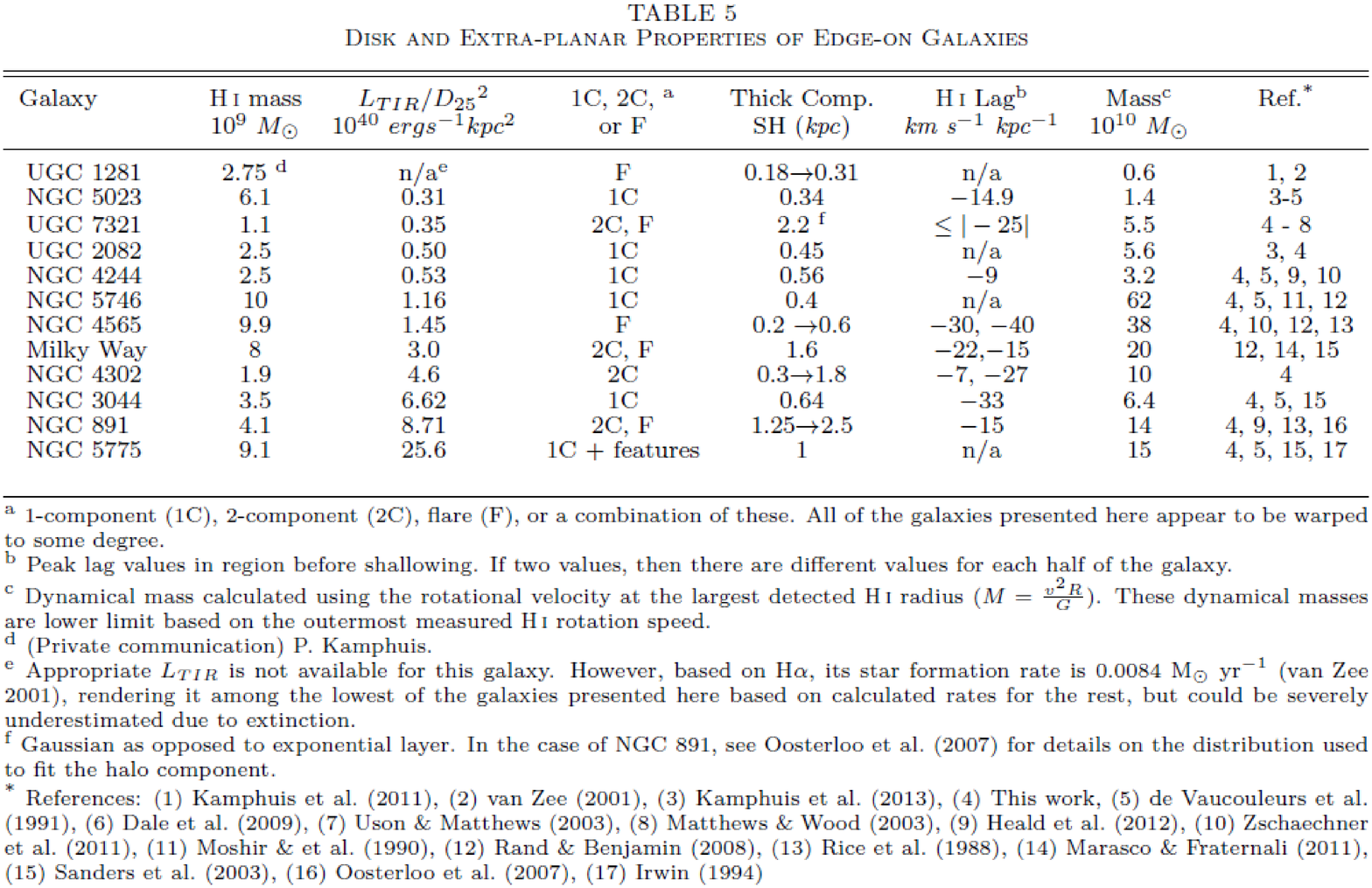}
 \label{table_5}
\end{figure*}

    We consider trends involving total infrared luminosity per unit area ($L_{TIR}$/${D_{25}}^2$), which is effectively a measure of SFR per unit area and is less affected by extinction than H$\alpha$. In Table~\ref{table_5} we compare properties of the galaxies presented here and throughout the literature, including total {\sc H\,i} mass, $L_{TIR}$/${D_{25}}^2$, scale heights, dynamical masses, and lags.  $L_{TIR}$/${D_{25}}^2$ calculations are done based on equations found in \citet{2009ApJ...703..517D}, and involve the optical area (\citealt{1991trcb.book.....D} in most cases; additional references are provided in the table).  Spitzer MIPS data are used if available for a given galaxy, otherwise IRAS data are used.

\subsection{Connections Between {\sc H\,i} Lags and Star Formation}\label{lags.sfr}

\par
   \citet{2007ApJ...663..933H} remarked on a possible connection between low global SFR and steeper DIG lags.  Additionally, in that work it was suggested that lag magnitude per DIG scale height might be a constant value among galaxies.  However, that work involved only three galaxies: NGC 891, NGC 5775, and NGC 4302, rendering trends difficult to constrain at that time.  Of the galaxies considered in this work, only NGC 4302 has a measured and modeled DIG lag.  However, the analysis of a larger sample of DIG lags, which can be compared with {\sc H\,i} lags, is currently underway by another HALOGAS team member and results are pending (e.g.\ \citealt{2013AAS...22114625W}).

\par
    For this analysis, we consider {\sc H\,i} lags interior to the radius where they begin to shallow.  These are the values listed in Table~\ref{table_5}.  As can be seen by examination in Table~\ref{table_5}, there appears to be no such inverse relation between {\sc H\,i} lag magnitude and $L_{TIR}$/${D_{25}}^2$ within a given galaxy.  In fact, the shallowest lag we measure is in NGC 4244, the galaxy having the lowest $L_{TIR}$/${D_{25}}^2$ among our sample.  In contrast, NGC 3044 has a much steeper lag, and a much higher $L_{TIR}$/${D_{25}}^2$.  Thus, the {\sc H\,i} lags clearly do not follow the \citet{2007ApJ...663..933H} trend.  This lack of a correlation is also seen when considering lag magnitudes per scale height of the thick component.  In the ballistic models of \citet{2002ApJ...578...98C}, the lag grows with the ratio of the vertical launch velocity of clouds from the disk to the rotation speed, the latter reflecting the ability of gravity to decelerate the flow.  So, to approximately remove the effect of gravity, we multiply the lags by the average rotation speed (for the flat part of the rotation curve) and still find no correlation.

\subsection{Radial Variations of Observed {\sc H\,i} Lags}\label{radialvariations}
\par
\begin{figure}
\centering
\includegraphics[width=80mm]{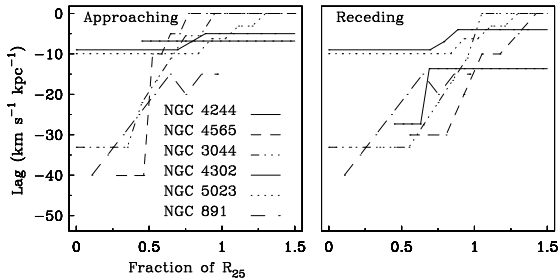}
\caption{\scriptsize\textit{The radial variations of the optimal {\sc H\,i} lags modeled for NGC 4244, NGC 4565, NGC 3044, NGC 4302, NGC 5023, and NGC 891.  The radial distance in each galaxy is normalized to R$_{25}$.  Note how all of the lags reach their shallowest values near R$_{25}$.} \label{trendsparameterlag}}
\end{figure}

    The lags in NGC 3044 and NGC 4302 appear to shallow radially (although only on the receding half of NGC 4302 and the complicated nature of that galaxy renders such determination difficult and skepticism is warranted in that case).  This is also true of the lags of NGC 891 \citep{2007AJ....134.1019O}, NGC 4244 \citep{2011ApJ...740...35Z}, NGC 4565 \citep{2012ApJ...760...37Z}, and possibly NGC 5023 \citep{2013MNRAS.434.2069K}.  These lags decrease in magnitude with radius, shallowing between approximately 0.5 and 1.0 times $R_{25}$ (Figure~\ref{trendsparameterlag}).

\section{Implications for Theoretical Scenarios}\label{theory}
\par
    The magnitudes of the lags are clearly steeper than those produced by ballistic models (e.g.\ \citealt{2002ApJ...578...98C}).  Purely ballistic models have already been known to underestimate lags.  One possible reason for this discrepancy is angular momentum lost from material ejected high above the plane of the disk in disk-halo flows to a slow-rotating, hot halo encompassing a galaxy as discussed in $\S$~\ref{introduction} \citep{2011MNRAS.415.1534M}.  

\par
   In the context of the models of \citet{2011MNRAS.415.1534M} in which hot coronal gas is mixed with galactic fountain flows, cools, and then is accreted onto the disk, then the radial shallowing of the observed {\sc H\,i} lags indicates that a majority of this accretion must take place within a certain radius near $R_{25}$. Additionally, with the possible exception of NGC 4302, the lags all start close to the midplane.  The primary implication of the lags occurring within R$_{25}$ and starting at a low z is that it is unlikely that they are purely accretion driven, and disk-halo flows are likely largely responsible for them.

\par
   The lags \citet{2011MNRAS.415.1534M} simulate are approximately $-$15 km s$^{-1}$ kpc$^{-1}$.  These values match the lags in NGC 891 and the Milky Way, but the lags we measure in NGC 4565 and NGC 3044 are steeper by a factor of two.  However, this could merely be due to a difference in how the lags are quantified.  In a radially varying model presented in \citet{2007AJ....134.1019O}, the lag in NGC 891 is closer to $-$30 km s$^{-1}$ kpc$^{-1}$ near the center.

\par
  That the radial shallowing of the lags occurs over roughly the same radial range (in that the lags reach their shallowest values near R$_{25}$) for the galaxies in Figure~\ref{trendsparameterlag} suggests that some internal process is largely responsible.  As discussed in $\S$~\ref{introduction}, some shallowing is expected from considerations of conservation of angular momentum, but the overall steepness of lags and the degree of shallowing suggests other physics at work in disk-halo flow dynamics.  Recall the ideas put forth by Benjamin (2002, 2012) in which extra-planar pressure gradients could modify lags.  Such considerations could be used to constrain the overall lag magnitudes as well as their radial variations.  Currently there are few observational constraints for such pressure gradients.  This situation may soon change, e.g.\, it should be possible to estimate extra-planar non-thermal radial pressure gradients via the deep continuum observations of edge-on galaxies described by \citet{2012AJ....144...43I}.  Almost all of the galaxies in Table~\ref{table_5} are in their sample.

\section{Summary \& Conclusions} 
\subsection{NGC 3044}

\par
   NGC 3044 is a somewhat lop-sided galaxy with indications of having undergone a disturbance at some point.  It is a relatively isolated galaxy, but that does not rule out past merger activity. There are several distinct {\sc H\,i} features within NGC 3044, some of which are extra-planar.     

\par
   Through modeling of the {\sc H\,i} distribution and kinematics, NGC 3044 is found to be comprised of a single, thick disk with a scale height of 7" (635 pc).  This disk has an inclination of 85$^{\circ}$, and is somewhat warped across the line of sight. The rotation curves on the two halves of the galaxy are either asymmetric by roughly 20 km s$^{-1}$, or else the inner and outer regions of the galaxy have differing systemic velocities.   We detect a relatively steep lag of $-$33$^{+6}_{-11}$ km s$^{-1}$ kpc$^{-1}$ on the approaching side and $-$33$\pm{6}$ km s$^{-1}$ kpc$^{-1}$ on the receding that shallows on both sides with radius, vanishing near $R_{25}$.  

\subsection{NGC 4302}

\par
   We present deep {\sc H\,i} observations and detailed models of NGC 4302.  We find the best fit model to consist of a thin disk with a scale height of 4" (300 pc), and a second, thicker disk with a scale height of 25" (1.8 kpc) that begins at a radius of 1.75' (7.7 kpc).  To this model we add a component to model the ram pressure stripped gas, as well as two additional features, one possibly due to spiral structure, while the other may be related to tidal interactions, accretion from NGC 4298, or disk-halo flows. 

\par
   Our optimal model includes a lag of $-$7$^{+3}_{-7}$ km s$^{-1}$ kpc$^{-1}$ on the approaching half, and $-$27$^{+21}_{-55}$ km s$^{-1}$ kpc$^{-1}$ on the receding half.

\subsection{Summary of Lag Trends}

\par
   We consider lags in 12 edge-on galaxies and there appears to be no correlation between lag magnitude and our measure of SFR per unit disk area as was suggested by the DIG lags presented in \citet{2007ApJ...663..933H}.  {\sc H\,i} lags show a great range.  The comparative steepness of the lags in NGC 3044 and NGC 4565 compared to NGC 4244 again emphasizes that lags differ greatly between galaxies, which must be considered in future simulations if lag origins are to be understood. 

\par
   Lags now have been observed to shallow with radius in six galaxies (although in the case of NGC 4302 this is only seen in the receding half, and all lag results for that galaxy should be considered skeptically).  This shallowing occurs between approximately 0.5 and 1.0 times $R_{25}$, suggesting internal dynamical processes are important in determining lag behavior.  Theoretical simulations have yet to incorporate such shallowing, which will be a valuable constraint in the future.

\section{Acknowledgments}

\par 
     We thank the operators at VLA for overseeing the observations.  The National Radio Astronomy Observatory is a facility of the National 
Science Foundation operated under cooperative agreement by Associated Universities, Inc.  This research has made use of the NASA/IPAC Infrared Science Archive, which is operated by the Jet Propulsion Laboratory, California Institute of Technology, under contract with the National Aeronautics and Space Administration.  We also thank Gregory B. Taylor and Peter Zimmer for constructive comments.  This material is based on work partially supported by the National Science Foundation under grant AST-0908106 to R.J.R. and R.W.  Finally, we thank the anonymous referee for insightful and constructive comments leading to improvement of this manuscript.

\bibliographystyle{apj}
\bibliography{3044_4302_draft.bib}

\begin{thebibliography}{52}
\expandafter\ifx\csname natexlab\endcsname\relax\def\natexlab#1{#1}\fi

\bibitem[{{Benjamin}(2002)}]{2002ASPC..276..201B}
{Benjamin}, R.~A. 2002, in Astronomical Society of the Pacific Conference
  Series, Vol. 276, Seeing Through the Dust: The Detection of HI and the
  Exploration of the ISM in Galaxies, ed. {A.~R.~Taylor, T.~L.~Landecker, \&
  A.~G.~Willis}, 201--+

\bibitem[{{Benjamin}(2012)}]{2012EAS....56..299B}
{Benjamin}, R.~A. 2012, in EAS Publications Series, Vol.~56, EAS Publications
  Series, ed. M.~A. {de Avillez}, 299--304

\bibitem[{{Boomsma} {et~al.}(2005){Boomsma}, {Oosterloo}, {Fraternali}, {van
  der Hulst}, \& {Sancisi}}]{2005A&A...431...65B}
{Boomsma}, R., {Oosterloo}, T.~A., {Fraternali}, F., {van der Hulst}, J.~M., \&
  {Sancisi}, R. 2005, \aap, 431, 65

\bibitem[{{Caimmi}(2008)}]{2008NewA...13..314C}
{Caimmi}, R. 2008, NA, 13, 314 

\bibitem[{{Chung} {et~al.}(2009){Chung}, {van Gorkom}, {Kenney}, {Crowl}, \&
  {Vollmer}}]{2009AJ....138.1741C}
{Chung}, A., {van Gorkom}, J.~H., {Kenney}, J.~D.~P., {Crowl}, H., \&
  {Vollmer}, B. 2009, \aj, 138, 1741

\bibitem[{{Chung} {et~al.}(2007){Chung}, {van Gorkom}, {Kenney}, \&
  {Vollmer}}]{2007ApJ...659L.115C}
{Chung}, A., {van Gorkom}, J.~H., {Kenney}, J.~D.~P., \& {Vollmer}, B. 2007,
  \apjl, 659, L115

\bibitem[{{Collins} {et~al.}(2002){Collins}, {Benjamin}, \&
  {Rand}}]{2002ApJ...578...98C}
{Collins}, J.~A., {Benjamin}, R.~A., \& {Rand}, R.~J. 2002, \apj, 578, 98

\bibitem[{{Collins} {et~al.}(2000){Collins}, {Rand}, {Duric}, \&
  {Walterbos}}]{2000ApJ...536..645C}
{Collins}, J.~A., {Rand}, R.~J., {Duric}, N., \& {Walterbos}, R.~A.~M. 2000,
  \apj, 536, 645

\bibitem[{{Dale} {et~al.}(2009){Dale}, {Cohen}, {Johnson}, {Schuster},
  {Calzetti}, {Engelbracht}, {Gil de Paz}, {Kennicutt}, {Lee}, {Begum},
  {Block}, {Dalcanton}, {Funes}, {Gordon}, {Johnson}, {Marble}, {Sakai},
  {Skillman}, {van Zee}, {Walter}, {Weisz}, {Williams}, {Wu}, \&
  {Wu}}]{2009ApJ...703..517D}
{Dale}, D.~A., {et~al.} 2009, \apj, 703, 517

\bibitem[{{de Vaucouleurs} {et~al.}(1991){de Vaucouleurs}, {de Vaucouleurs},
  {Corwin}, {Buta}, {Paturel}, \& {Fouque}}]{1991trcb.book.....D}
{de Vaucouleurs}, G., {de Vaucouleurs}, A., {Corwin}, Jr., H.~G., {Buta},
  R.~J., {Paturel}, G., \& {Fouque}, P. 1991, {Third Reference Catalogue of
  Bright Galaxies}, ed. {de Vaucouleurs, G., de Vaucouleurs, A., Corwin, H.~G.,
  Jr., Buta, R.~J., Paturel, G., \& Fouque, P.}

\bibitem[{{Fraternali} \& {Binney}(2006)}]{2006MNRAS.366..449F}
{Fraternali}, F., \& {Binney}, J.~J. 2006, \mnras, 366, 449

\bibitem[{{Fraternali} \& {Binney}(2008)}]{2008MNRAS.386..935F}
---. 2008, \mnras, 386, 935

\bibitem[{{Giovanelli} {et~al.}(2007){Giovanelli}, {Haynes}, {Kent},
  {Saintonge}, {Stierwalt}, {Altaf}, {Balonek}, {Brosch}, {Brown}, {Catinella},
  {Furniss}, {Goldstein}, {Hoffman}, {Koopmann}, {Kornreich}, {Mahmood},
  {Martin}, {Masters}, {Mitschang}, {Momjian}, {Nair}, {Rosenberg}, \&
  {Walsh}}]{2007AJ....133.2569G}
{Giovanelli}, R., {et~al.} 2007, \aj, 133, 2569

\bibitem[{{Heald} {et~al.}(2011){Heald}, {J{\'o}zsa}, {Serra}, {Zschaechner},
  {Rand}, {Fraternali}, {Oosterloo}, {Walterbos}, {J{\"u}tte}, \&
  {Gentile}}]{2011A&A...526A.118H}
{Heald}, G., {et~al.} 2011, \aap, 526, A118+

\bibitem[{{Heald} {et~al.}(2012){Heald}, {J{\'o}zsa}, {Serra}, {Zschaechner},
  {Rand}, {Fraternali}, {Oosterloo}, {Walterbos}, {J{\"u}tte}, \&
  {Gentile}}]{2012A&A...544C...1H}
---. 2012, \aap, 544, C1

\bibitem[{{Heald} {et~al.}(2007){Heald}, {Rand}, {Benjamin}, \&
  {Bershady}}]{2007ApJ...663..933H}
{Heald}, G.~H., {Rand}, R.~J., {Benjamin}, R.~A., \& {Bershady}, M.~A. 2007,
  \apj, 663, 933

\bibitem[{{Hodges-Kluck} \& {Bregman}(2013)}]{2013ApJ...762...12H}
{Hodges-Kluck}, E.~J., \& {Bregman}, J.~N. 2013, \apj, 762, 12

\bibitem[{{Howk} \& {Savage}(1999)}]{1999AJ....117.2077H}
{Howk}, J.~C., \& {Savage}, B.~D. 1999, \aj, 117, 2077

\bibitem[{{Irwin} {et~al.}(2012){Irwin}, {Beck}, {Benjamin}, {Dettmar},
  {English}, {Heald}, {Henriksen}, {Johnson}, {Krause}, {Li}, {Miskolczi},
  {Mora}, {Murphy}, {Oosterloo}, {Porter}, {Rand}, {Saikia}, {Schmidt},
  {Strong}, {Walterbos}, {Wang}, \& {Wiegert}}]{2012AJ....144...43I}
{Irwin}, J., {et~al.} 2012, \aj, 144, 43

\bibitem[{{Irwin}(1994)}]{1994ApJ...429..618I}
{Irwin}, J.~A. 1994, \apj, 429, 618

\bibitem[{{Irwin} {et~al.}(2013){Irwin}, {Brar}, {Saikia}, \&
  {Henriksen}}]{2013MNRAS.433.2958I}
{Irwin}, J.~A., {Brar}, R.~S., {Saikia}, D.~J., \& {Henriksen}, R.~N. 2013,
  \mnras, 433, 2958

\bibitem[{{Irwin} {et~al.}(1999){Irwin}, {English}, \&
  {Sorathia}}]{1999AJ....117.2102I}
{Irwin}, J.~A., {English}, J., \& {Sorathia}, B. 1999, \aj, 117, 2102

\bibitem[{{J{\'o}zsa} {et~al.}(2007){J{\'o}zsa}, {Kenn}, {Klein}, \&
  {Oosterloo}}]{2007A&A...468..731J}
{J{\'o}zsa}, G.~I.~G., {Kenn}, F., {Klein}, U., \& {Oosterloo}, T.~A. 2007,
  \aap, 468, 731

\bibitem[{{Kamphuis} {et~al.}(2011){Kamphuis}, {Peletier}, {van der Kruit}, \&
  {Heald}}]{2011MNRAS.414.3444K}
{Kamphuis}, P., {Peletier}, R.~F., {van der Kruit}, P.~C., \& {Heald}, G.~H.
  2011, \mnras, 414, 3444

\bibitem[{{Kamphuis} {et~al.}(2013){Kamphuis}, {Rand}, {J{\'o}zsa},
  {Zschaechner}, {Heald}, {Patterson}, {Gentile}, {Walterbos}, {Serra}, \& {de
  Blok}}]{2013MNRAS.434.2069K}
{Kamphuis}, P., {et~al.} 2013, \mnras, 434, 2069

\bibitem[{{Lee} \& {Irwin}(1997)}]{1997ApJ...490..247L}
{Lee}, S.-W., \& {Irwin}, J.~A. 1997, \apj, 490, 247

\bibitem[{{Marasco} \& {Fraternali}(2011)}]{2011A&A...525A.134M}
{Marasco}, A., \& {Fraternali}, F. 2011, \aap, 525, A134

\bibitem[{{Marinacci} {et~al.}(2011){Marinacci}, {Fraternali}, {Nipoti},
  {Binney}, {Ciotti}, \& {Londrillo}}]{2011MNRAS.415.1534M}
{Marinacci}, F., {Fraternali}, F., {Nipoti}, C., {Binney}, J., {Ciotti}, L., \&
  {Londrillo}, P. 2011, \mnras, 415, 1534

\bibitem[{{Matthews} \& {Uson}(2008)}]{2008ApJ...688..237M}
{Matthews}, L.~D., \& {Uson}, J.~M. 2008, \apj, 688, 237

\bibitem[{{Matthews} \& {Wood}(2003)}]{2003ApJ...593..721M}
{Matthews}, L.~D., \& {Wood}, K. 2003, \apj, 593, 721

\bibitem[{{Moshir} \& {et al.}(1990)}]{1990IRASF.C......0M}
{Moshir}, M., \& {et al.} 1990, in IRAS Faint Source Catalogue, version 2.0
  (1990), 0

\bibitem[{{Oosterloo} {et~al.}(2007){Oosterloo}, {Fraternali}, \&
  {Sancisi}}]{2007AJ....134.1019O}
{Oosterloo}, T., {Fraternali}, F., \& {Sancisi}, R. 2007, \aj, 134, 1019

\bibitem[{{Putman} {et~al.}(2012){Putman}, {Peek}, \&
  {Joung}}]{2012ARA&A..50..491P}
{Putman}, M.~E., {Peek}, J.~E.~G., \& {Joung}, M.~R. 2012, \araa, 50, 491

\bibitem[{{Rand}(1996)}]{1996ApJ...462..712R}
{Rand}, R.~J. 1996, \apj, 462, 712

\bibitem[{{Rand} \& {Benjamin}(2008)}]{2008ApJ...676..991R}
{Rand}, R.~J., \& {Benjamin}, R.~A. 2008, \apj, 676, 991

\bibitem[{{Rice} {et~al.}(1988){Rice}, {Lonsdale}, {Soifer}, {Neugebauer},
  {Kopan}, {Lloyd}, {de Jong}, \& {Habing}}]{1988ApJS...68...91R}
{Rice}, W., {Lonsdale}, C.~J., {Soifer}, B.~T., {Neugebauer}, G., {Kopan},
  E.~L., {Lloyd}, L.~A., {de Jong}, T., \& {Habing}, H.~J. 1988, \apjs, 68, 91

\bibitem[{{Rossa} \& {Dettmar}(2003)}]{2003A&A...406..493R}
{Rossa}, J., \& {Dettmar}, R. 2003, \aap, 406, 493

\bibitem[{{Rueff} {et~al.}(2013){Rueff}, {Howk}, {Pitterle}, {Hirschauer},
  {Fox}, \& {Savage}}]{2013AJ....145...62R}
{Rueff}, K.~M., {Howk}, J.~C., {Pitterle}, M., {Hirschauer}, A.~S., {Fox},
  A.~J., \& {Savage}, B.~D. 2013, \aj, 145, 62

\bibitem[{{Sancisi} {et~al.}(2008){Sancisi}, {Fraternali}, {Oosterloo}, \& {van
  der Hulst}}]{2008A&ARv..15..189S}
{Sancisi}, R., {Fraternali}, F., {Oosterloo}, T., \& {van der Hulst}, T. 2008,
  \aapr, 15, 189

\bibitem[{{Sanders} {et~al.}(2003){Sanders}, {Mazzarella}, {Kim}, {Surace}, \&
  {Soifer}}]{2003AJ....126.1607S}
{Sanders}, D.~B., {Mazzarella}, J.~M., {Kim}, D.-C., {Surace}, J.~A., \&
  {Soifer}, B.~T. 2003, \aj, 126, 1607

\bibitem[{{Solomon} \& {Sage}(1988)}]{1988ApJ...334..613S}
{Solomon}, P.~M., \& {Sage}, L.~J. 1988, \apj, 334, 613

\bibitem[{{Sommer-Larsen} {et~al.}(2003){Sommer-Larsen}, {G{\"o}tz}, \&
  {Portinari}}]{2003ApJ...596...47S}
{Sommer-Larsen}, J., {G{\"o}tz}, M., \& {Portinari}, L. 2003, \apj, 596, 47

\bibitem[{{Springob} {et~al.}(2005){Springob}, {Haynes}, {Giovanelli}, \&
  {Kent}}]{2005ApJS..160..149S}
{Springob}, C.~M., {Haynes}, M.~P., {Giovanelli}, R., \& {Kent}, B.~R. 2005,
  \apjs, 160, 149

\bibitem[{{Swaters} {et~al.}(1997){Swaters}, {Sancisi}, \& {van der
  Hulst}}]{1997ApJ...491..140S}
{Swaters}, R.~A., {Sancisi}, R., \& {van der Hulst}, J.~M. 1997, \apj, 491, 140

\bibitem[{{T{\"u}llmann} {et~al.}(2006){T{\"u}llmann}, {Pietsch}, {Rossa},
  {Breitschwerdt}, \& {Dettmar}}]{2006A&A...448...43T}
{T{\"u}llmann}, R., {Pietsch}, W., {Rossa}, J., {Breitschwerdt}, D., \&
  {Dettmar}, R. 2006, \aap, 448, 43

\bibitem[{{Tully} {et~al.}(1988){Tully}, {Fisher}, \&
  {Madore}}]{1988JRASC..82..305T}
{Tully}, R.~B., {Fisher}, J.~R., \& {Madore}, B.~F. 1988, \jrasc, 82, 305

\bibitem[{{Uson} \& {Matthews}(2003)}]{2003AJ....125.2455U}
{Uson}, J.~M., \& {Matthews}, L.~D. 2003, \aj, 125, 2455

\bibitem[{{van Zee}(2001)}]{2001AJ....121.2003V}
{van Zee}, L. 2001, \aj, 121, 2003

\bibitem[{{Wakker} \& {van Woerden}(1997)}]{1997ARA&A..35..217W}
{Wakker}, B.~P., \& {van Woerden}, H. 1997, \araa, 35, 217

\bibitem[{{Wu} {et~al.}(2013){Wu}, {Walterbos}, {Rand}, {Heald}, \&
  {HALOGAS}}]{2013AAS...22114625W}
{Wu}, C.~J., {Walterbos}, R.~A., {Rand}, R.~J., {Heald}, G., \& {HALOGAS}.
  2013, in American Astronomical Society Meeting Abstracts, Vol. 221, American
  Astronomical Society Meeting Abstracts, 146.25

\bibitem[{{Zschaechner} {et~al.}(2012){Zschaechner}, {Rand}, {Heald},
  {Gentile}, \& {J{\'o}zsa}}]{2012ApJ...760...37Z}
{Zschaechner}, L.~K., {Rand}, R.~J., {Heald}, G.~H., {Gentile}, G., \&
  {J{\'o}zsa}, G. 2012, \apj, 760, 37

\bibitem[{{Zschaechner} {et~al.}(2011){Zschaechner}, {Rand}, {Heald},
  {Gentile}, \& {Kamphuis}}]{2011ApJ...740...35Z}
{Zschaechner}, L.~K., {Rand}, R.~J., {Heald}, G.~H., {Gentile}, G., \&
  {Kamphuis}, P. 2011, \apj, 740, 35

\end{thebibliography}

\nocite{2002ASPC..276..201B}
\nocite{2012EAS....56..299B}

\end{document}